\documentclass[letterpaper,onecolumn,nofootinbib,11pt]{revtex4-1}
\pdfoutput=1

\newcommand{\dropcap}[1]{#1}

\usepackage[margin=2cm]{geometry}
\usepackage{amsmath}
\usepackage{amssymb}
\usepackage{graphicx}
\usepackage{textcomp}
\usepackage{hyperref}
\usepackage{graphics}
\usepackage{pbox}
\usepackage{graphicx}
\usepackage{xcolor} 
\usepackage{tikz}
\usepackage{rotating}

\usetikzlibrary{decorations.pathreplacing}

\begin{document}

\title{Experimental demonstration of an isotope-sensitive warhead verification technique using nuclear resonance fluorescence}

\author{Jayson R.~Vavrek}\affiliation{Laboratory for Nuclear Security and Policy, Massachusetts Institute of 
Technology, Cambridge, MA 02139}
  
\author{Brian S.~Henderson}\affiliation{Laboratory for Nuclear Security and Policy, Massachusetts Institute of 
Technology, Cambridge, MA 02139}
  
\author{Areg Danagoulian}\affiliation{Laboratory for Nuclear Security and Policy, Massachusetts Institute of 
Technology, Cambridge, MA 02139}

\date{\today}

\begin{abstract}
 Future nuclear arms reduction efforts will require technologies to verify that 
warheads slated for dismantlement are authentic without revealing any sensitive 
weapons design information to international inspectors. Despite several decades 
of research, no technology has met these requirements simultaneously. Recent 
work by Kemp et al. [Kemp RS, Danagoulian A, Macdonald RR, Vavrek JR (2016) Proc Natl Acad Sci USA 113:8618--8623] has produced a novel physical cryptographic verification protocol that approaches this treaty verification problem by 
exploiting the isotope-specific nature of nuclear resonance fluorescence (NRF) 
measurements to verify the authenticity of a warhead. To 
protect sensitive information, the NRF signal from the warhead is convolved 
with that of an encryption foil that contains key warhead isotopes in amounts
unknown to the inspector. The convolved spectrum from a candidate warhead is 
statistically compared against that from an authenticated template warhead to 
determine whether the candidate itself is authentic. Here we report on recent 
proof-of-concept warhead verification experiments conducted at the Massachusetts Institute of Technology. Using 
high-purity germanium (HPGe) detectors, we measured NRF spectra from the 
interrogation of proxy `genuine' and `hoax' objects by a 2.52~MeV endpoint
bremsstrahlung beam. The observed differences in NRF intensities near 2.2~MeV 
indicate that the physical cryptographic protocol can distinguish between proxy 
genuine and hoax objects with high confidence in realistic measurement times.
\end{abstract}


\maketitle



\dropcap{N}uclear arms reduction treaties have traditionally suffered from the
disarmament verification problem: how can one confidently identify a warhead as 
authentic without having access to any sensitive design information that proves 
it is authentic? Rather than confront this apparent paradox, treaties such as 
the New Strategic Arms Reduction Treaty (New START) have relied on verification 
of warhead delivery vehicles---e.g.,~missiles and bomber aircraft---rather than 
direct verification of warheads themselves. Future arms control agreements, 
however, may require some mechanism for the verification of individual 
warheads~\cite{drell1993verification,comley2000confidence,spears2001technology,
fuller}
to ensure that a country does not dispose of fraudulent or `hoax' warheads in a 
gambit to obtain a strategic nuclear advantage.

\section*{Warhead verification}
In a warhead verification protocol, a warhead owner (`host') attempts to prove 
to an inspection team (`inspector') that an object submitted for inspection and 
subsequent dismantlement and disposition is indeed a genuine nuclear warhead. 
An 
object successfully verified may then be dismantled by the host 
under a secure
chain of custody~\cite{bunchCoC} and counted 
towards the host's obligations under an arms reduction treaty. At the same 
time, 
the host seeks to prevent the inspector from learning any sensitive information 
about the design of the warhead, whether to prevent proliferation of nuclear 
weapons technology or disclosure of warhead architecture and vulnerabilities. 
Thus, the verification measurement must be designed and performed in such a 
way as to provide a strong test of authenticity while while minimizing 
intrusiveness and maximizing information security. Non-authentic warheads 
(`hoaxes') fall into two broad categories: isotopic hoaxes, in which a valuable 
weapon component (e.g.,~the weapons-grade Pu fissile fuel) is replaced by a 
less-valuable surrogate of similar geometry (e.g.,~reactor grade Pu); and 
geometric hoaxes, in which isotopes are present in their correct amounts but in 
a non-weapons-usable configuration (e.g.,~rough slabs of Pu rather than 
highly-engineered spherical shells).

Past approaches to warhead verification have generally focused on the 
`attribute' approach, in which the protocol measures a set of key
characteristics thought to define a warhead, such as the total mass of 
plutonium 
and the isotopic ratio of Pu-239 to Pu-240 in the object~\cite{lanl2001fmttd}. 
Such measurements are highly intrusive, and so are conducted behind an 
`information barrier' (IB), an electronic or software layer that shields the 
classified raw measurement data 
and presents the inspector with only a binary pass/fail answer for each of the 
attribute measurements~\cite{close2001infobarriers}. However, certifying that an 
electronic or (especially) software IB does not contain any hidden backdoors or 
functionalities---which a 
nefarious inspector could exploit to obtain sensitive information or a 
nefarious 
host could use to fraudulently simulate a `genuine' result---is exceedingly 
difficult, and may never be satisfactorily proven. Moreover, attributes must be 
chosen specifically to describe real nuclear warheads, and thus may constitute 
sensitive information themselves. Even then, the set of attributes may not be 
complete, opening the door to hoax objects that pass all the attribute tests 
but 
nevertheless are not real warheads.

More recent work has therefore focused on the `template' approach to 
verification, in which comparison to a known genuine object (the ``template'') 
is used to certify subsequent objects presented for 
inspection~\cite{yan2015review,fuller,marleau2015tabletop}. In such a 
protocol, the measurements of both the template and subsequent objects are 
encrypted using the same method, so that only the encrypted signals (or 
``hashes'') must be compared to authenticate. The hash should be unique to a 
particular combination of geometry and isotopic makeup (i.e.,~a particular 
warhead design), while containing no sensitive information about the object. As 
such, the hash is useless on its own, and only has any use in comparison 
against 
the hash of another object---a warhead that is already known to be genuine. 
This 
authenticated template warhead could be established for instance via an 
unannounced visit by the inspector to a random launch facility in the host
country, and then by selecting a random warhead from an active-duty 
intercontinental ballistic missile.\footnote{In any template warhead 
verification protocol, the utility of every measurement hinges on the 
authenticity of the template. A complete solution to the question of first 
establishing such an authentic template will require classified knowledge of the 
chain of custody of a country's nuclear stockpile, and therefore is an open question beyond the scope of this article.} A measurement of the 
authenticated template would then be used as the standard against which to 
compare the measurements from the same model of warhead covered by the arms 
control treaty.

Recent papers have put forth template verification protocols that aim to make a 
verification measurement of a warhead while protecting sensitive design 
information. A team of researchers at Princeton proposed and later 
experimentally demonstrated a verification protocol using superheated bubble 
detectors and fast neutron radiography~\cite{ref:alex,philippe2016verification}.
In parallel, a team at the Massachusetts Institute of Technology (MIT) developed an alternative approach using isotopic 
tomography via transmission nuclear resonance fluorescence 
(tNRF)~\cite{kemp2016physical}; the present work is an experimental 
demonstration of the MIT tNRF protocol. Further techniques using 
coded-aperture-based passive neutron counters~\cite{marleau2017implementations}
and epithermal neutron resonance radiography \cite{hecla2017epithermal}, from 
Sandia National Laboratories and MIT, respectively, have been proposed in the 
past year.

The strengths and weaknesses of the aforementioned proposals can be compared by 
examining the three requirements of an ideal warhead verification protocol:
\begin{enumerate}
    \item completeness: the ideal protocol must clear all real warheads;
    \item soundness: the ideal protocol must raise an alarm on all hoax 
warheads;
    \item information security: the ideal protocol must be 
\textit{zero-knowledge}~\cite{goldwasser1989knowledge,blum1988zk}---for an 
honest host, it must not reveal anything beyond a binary genuine/hoax 
determination.
\end{enumerate}
The Princeton protocol is essentially zero-knowledge, returning a flat image 
(up 
to statistical variation) if the host has submitted a real warhead. In its 
original form~\cite{ref:alex}, the measurement faces a challenge in the 
soundness requirement: fast neutron radiography is insensitive to the isotopic 
or (in some cases) elemental composition of the object, and cannot on its own 
distinguish between weapon materials and well-chosen hoax materials. Additional 
measurement modes using multiple incident neutron energies~\cite{yan2015two} 
have been proposed to increase the protocol's discrimination between fissionable 
and fissile isotopes. Similarly, work on the Sandia coded-aperture protocol has 
focused on satisfying the completeness and information security aspects of the 
problem, but has not demonstrated resistance to hoaxing by a neutron source of 
similar geometry and activity.

Unlike the Princeton and Sandia protocols, the two MIT protocols are highly 
sensitive to isotopics through their use of isotope-specific resonant 
phenomena, 
making them highly robust against a large class of hoaxes. While the MIT tNRF 
protocol is not zero-knowledge (since the inspector has access to the hashed 
measurements rather than solely a binary genuine/hoax determination), and thus 
there are uncertainties about the extent of the information security of the MIT 
tNRF protocol, there are methods to make it sufficiently secure~\cite{kemp2016physical}. This work demonstrates the core measurement of the MIT 
tNRF protocol, and is an experimental implementation of an 
isotopically-sensitive warhead 
verification measurement.

\section*{Nuclear resonance fluorescence measurements}
Nuclear resonance fluorescence (NRF) describes the X$(\gamma, \gamma')$X 
reaction in which a photon $\gamma$ is resonantly absorbed by the nucleus X and 
then re-emitted as the excited nucleus subsequently transitions to its ground 
state~\cite{metzger1959resonance,kneissl1996structure}. The cross 
section for an NRF interaction with absorption via the resonant energy level 
$E_r$ is given by the Breit-Wigner distribution
\begin{align}\label{eq:sigmaNRFBW}
    \sigma^\text{NRF}_{r}(E) = \pi g_r \left( \frac{\hbar c}{E_r} \right)^2 
\frac{\Gamma_r \Gamma_{r,0}}{(E-E_r)^2 + (\Gamma_r/2)^2}
\end{align}
where $\Gamma_r$ is the width of the level at $E_r$, $\Gamma_{r,0}$ is the 
partial width for transitions between $E_r$ and the ground state, and $g_r$ is 
a 
statistical factor as described in SI Appendix~\S \ref{sec:si_nrf}. For high-$Z$ 
isotopes 
of interest, these fundamental widths are typically ${\sim}10$~meV but 
the effective width of the cross section is increased to 
${\sim}1$~eV through Doppler broadening by thermal motion 
of the target nuclei. Imperfect detector resolution further broadens the 
measurable NRF 
resolution to widths of ${\sim}1$~keV. Since the NRF lines of an isotope are 
still typically $>$10~keV apart, the set of resonance energies $E_r$ provides a 
resolvable, one-to-one map between measurement space and isotopic space.

The MIT verification protocol exploits the isotope-specific nature 
of NRF to make a template measurement of the mass and geometry of the isotopes 
of interest to the inspector. As discussed in the following section and 
illustrated in Fig.~\ref{fig:schematic}, the measurement uses a broad-spectrum 
bremsstrahlung photon source to irradiate the measurement object; NRF 
interactions in the object preferentially attenuate the photon flux at specific 
energies determined by the unique nuclear energy level structure of each isotope 
according to how much of the isotope is present in the warhead.
The remaining transmitted flux at these energies goes on to induce further NRF 
interactions in an encryption foil, leading to NRF emission into high-purity 
germanium (HPGe) photon detectors at an observed rate (SI Appendix Eq.~\ref{eq:d2ndEdOmega}) 
that has been reduced by the presence of the NRF isotope in the warhead. The 
hashed measurements required for the template verification protocol are thus the 
recorded spectra, since it is impossible to precisely determine the warhead 
composition (i.e.,~the thickness $D$ in SI Appendix Eqs.~\ref{eq:phi_t} and 
\ref{eq:d2ndEdOmega}) from the height of the NRF peaks in the observed spectrum 
without knowledge of the \textit{detailed} composition of the foil (i.e.,~the 
thickness $X$ in SI Appendix Eq.~\ref{eq:d2ndEdOmega}). The exact foil 
design is therefore decided by the host and kept secret from the inspector. The 
influence of the warhead composition on the height of the NRF peaks---and thus 
any sensitive warhead design information---is then said to be \textit{physically 
encrypted} by the foil.  This technique uses the laws of physics to mask sensitive
information, rather than electronic or computer-based information barriers, making it 
substantially more robust against tampering and hoaxing than previously proposed techniques \cite{close2001infobarriers}.
Although the detailed construction of the foil is kept 
secret from the inspector in order to maintain the encryption, the mere presence 
of certain characteristic NRF lines in the detected spectrum corresponds to the 
presence of certain isotopes in the encryption foil, a fact the inspector may 
use to validate the utility of a given foil without breaking the encryption. 
The foil may also be placed under joint custody of the host and 
inspector to ensure it has not been altered between the template and 
candidate measurements. As an additional layer of information security, the 
host may add optional `encryption plates' of warhead materials to the measured 
object so that even if precise inference about the measured object is possible, 
it is impossible to infer anything about the warhead alone.

To protect against geometric hoaxes, the MIT protocol includes measurements of template and candidate
warheads in random or multiple random orientations due to the difficulty for the
host to engineer a hoax warhead that could mimic the template signal successfully
along multiple projections.  To increase the information security of this protocol, each orientation may be paired
with a unique cryptographic foil to dilute the information content of the multiple measurements.
Ref.~\cite{kemp2016physical} discusses the required complexity
of such geometric hoaxes, which increases rapidly with number of projections measured.

\section*{Experimental design}\label{sec:experiment}
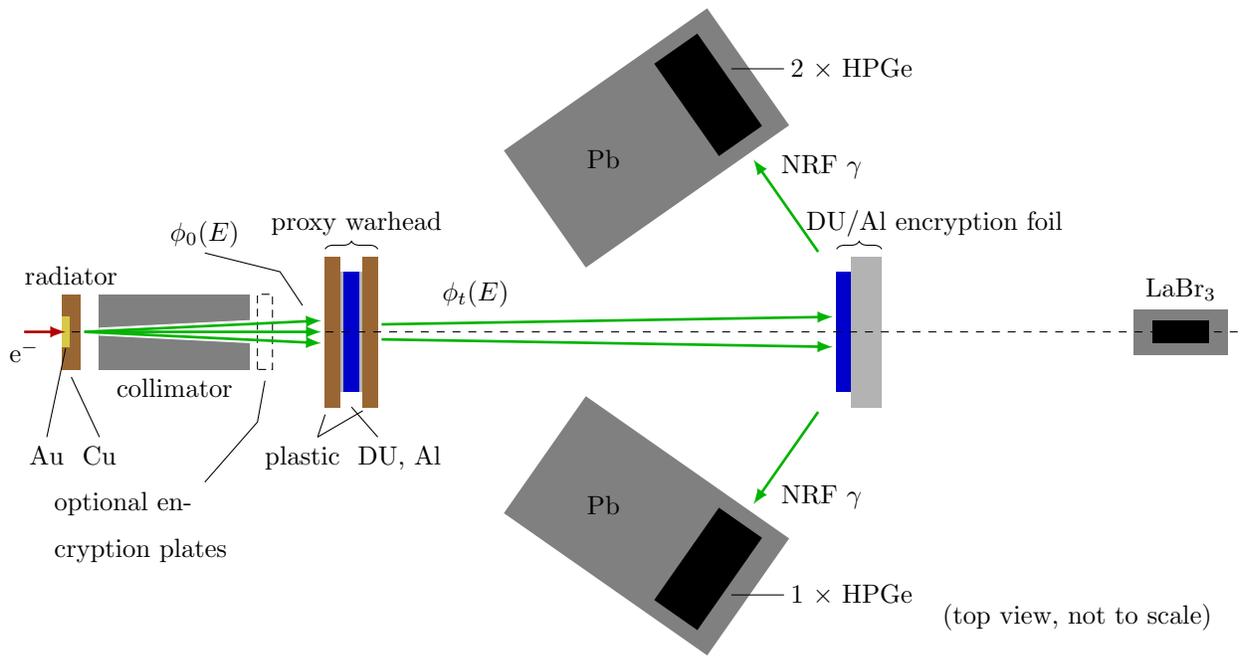
\begin{figure*}[ht]
\centering
\begin{tikzpicture}
\fill[fill=gray] (0,-0.5) rectangle (2,0.5);
\node[below] at (1,-0.5) {collimator};
\fill[fill=white, draw=white] (0,-0.05) -- (0,0.05) -- (2,0.15) -- (2,-0.15);

\draw[densely dashed] (2.1,-0.5) rectangle (2.3,0.5);
\draw (2.2, -0.65) -- (2.1, -1.2) -- (1.4, -2.0);
\node[below,nodes = {draw,align=center},text width=4cm] at (1.4, -2.0) {optional 
encryption plates};

\fill[fill=brown!80!black] (-0.5, -0.5) rectangle (-0.25, 0.5);
\fill[fill=yellow!80!black] (-0.5, -0.2) rectangle (-0.4, 0.2);
\node[above] at (-0.375,0.5) {radiator};
\draw (-0.375, -0.6) -- (0, -1.4);
\draw (-0.45, -0.22) -- (-0.7, -1.4);
\node[below] at (0,-1.4) {Cu};
\node[below] at (-0.7,-1.4) {Au};

\fill[fill=brown!80!black] (3,-1) rectangle (3.2,1);
\fill[fill=gray!60!white] (3.2,-0.8) rectangle (3.25,0.8); 
\fill[fill=blue!80!black] (3.25,-0.8) rectangle (3.45,0.8);
\draw [decorate,decoration={brace,amplitude=3pt}](3.0,1.1) -- (3.7,1.1) 
node[yshift=10pt,xshift=-8pt]{proxy warhead}; 
this
\fill[fill=gray!60!white] (3.45,-0.8) rectangle (3.5,0.8);
\fill[fill=brown!80!black] (3.5,-1) rectangle (3.7,1);
\node[below] at (2.7,-1.4) {plastic};
\node[below] at (3.99,-1.4) {DU, Al};
\draw (3.0,-1.1) -- (2.9,-1.4);
\draw (3.5,-1.05) -- (2.9,-1.4);
\draw (3.35,-1.) -- (3.7,-1.4);

\fill[fill=blue!80!black] (9.8,-0.8) rectangle (10,0.8);
\fill[fill=gray!60!white] (10,-1) rectangle (10.4,1);
\draw [decorate,decoration={brace,amplitude=3pt}]
(9.8,1.1) -- (10.4,1.1) node[xshift=20pt,yshift=10pt]
{DU/Al encryption foil};

\fill[fill=gray] (13.75,-0.30) rectangle (15.0,0.30);
\fill[fill=black] (14,-0.15) rectangle (14.75,0.15);
\node[above] at (14.375,0.30) {LaBr$_3$};

\fill[fill=gray, rotate around={55:(10.3,0)}] (5.5,-0.65) rectangle (7.4,2.65);
\fill[fill=gray, rotate around={-55:(10.3,0)}] (5.5,-2.65) rectangle (7.4,0.65);
\node at (6.7,2.3) {Pb};
\node at (6.7,-2.3) {Pb};

\fill[fill=black, rotate around={55:(10.3,0)}] (5.7,-0.35) rectangle (7.2,0.35);
\fill[fill=black, rotate around={-55:(10.3,0)}] (5.7,-0.35) rectangle 
(7.2,0.35);
\node at (10,3.5) {2 $\times$ HPGe};
\node at (10,-3.5) {1 $\times$ HPGe};
\draw (8.4,3.5) -- (9.1,3.5);
\draw (8.4,-3.5) -- (9.1,-3.5);

\draw[dashed] (-1,0) -- (15.2, 0);

\draw[-{latex}, green!70!black, line width=1pt] (-0.2,0) -- (2.95,  0.15);
\draw[-{latex}, green!70!black, line width=1pt] (-0.2,0) -- (2.95, -0.15);
\draw[-{latex}, green!70!black, line width=1pt] (-0.2,0) -- (2.95, 0);
\draw (2.7, 0.35) -- (2.4, 0.8) -- (1.4, 1.05);
\node[above] at (1.4, 1) {$\phi_0(E)$};

\draw[-{latex}, green!70!black, line width=1pt] (3.75, 0.1) -- (9.75, 0.2);
\draw[-{latex}, green!70!black, line width=1pt] (3.75, -0.1) -- (9.75, -0.2);
\node[above] at (5, 0.2) {$\phi_t(E)$};

\draw[-{latex}, green!70!black, line width=1pt, rotate around={-55:(10.3,0)}] 
(9, 0) -- (7.5, 0);
\draw[-{latex}, green!70!black, line width=1pt, rotate around={55:(10.3,0)}] 
(9, 
0) -- (7.5, 0);
\node at (9.6,2.2) {NRF $\gamma$};
\node at (9.6,-2.2) {NRF $\gamma$};

\draw[-{latex}, red!70!black, line width=1pt] (-1,0) -- (-0.45, 0);
\node[below] at (-1, 0) {e$^-$};

\node at (13.0, -3.8) {(top view, not to scale)};

\end{tikzpicture}
\caption{Schematic of the physical cryptographic NRF measurement. As an 
information security measure, the large Pb shields prevent the HPGe detectors 
from directly observing the the proxy warhead. Annotated photographs of the 
experiment geometry are shown in SI Appendix Figs.~\ref{fig:setup_photo} and 
\ref{fig:genuine_photo}.}
\label{fig:schematic}
\end{figure*}

Following the design depicted in Fig.~\ref{fig:schematic}, a bremsstrahlung
beam was used to illuminate a circular section of the object undergoing 
interrogation. Since no real nuclear warheads were available in an academic 
setting, several proxy warheads were constructed. The proxy warheads were 
objects with a set of isotopes---U-238 and Al-27---that form the basis for 
proof-of-concept NRF experiments and subsequent extrapolations to more 
realistic 
settings involving weapon isotopes such as U-235, Pu-239, and Pu-240. The first 
proxy genuine target (``template~I'') was constructed from DU plates of total 
thickness $3.18$~mm (wrapped in thin layers of Al foil, amounting to a total 
thickness of ${\sim}0.25$~mm) encased between two 
$19$~mm-thick layers of high-density polyethylene (HDPE) as proxy high 
explosives. In the first hoax target (``hoax~Ia''), the DU was replaced by 
$5.29$~mm of Pb sheets in order to match the nominal areal densities of high-$Z$ 
material to better than $1\%$. A second measurement of template I was made on 
the following day of experiments to emulate the verification of a genuine 
candidate warhead (``candidate~Ig''). The Pb hoax was similarly re-measured 
(``hoax Ib''). To emulate measurements on different warhead designs, a second 
genuine target (``template~II'') with double the thickness of DU was also 
tested 
against a hoax with double the thickness of Pb (``hoax~IIc'') and against a 
partial hoax (``hoax~IId'') in which only half the DU was replaced. In total, 
seven measurements were conducted on five different targets (see 
Table~\ref{tab:configs_small} and Figs.~\ref{fig:I_vs_Ia}--\ref{fig:II_vs_IId}).

Experiments were performed at MIT's High Voltage Research Laboratory (HVRL), 
which houses a continuous-wave Van de Graaff electron accelerator capable 
of producing electron kinetic energies of $2.0$--$3.0$~MeV at beam currents of 
up to $30$~{\textmu}A. For the physical cryptography measurements, a $2.52$~MeV
electron beam at the maximum stable current (between 25--30~{\textmu}A) was 
directed towards a water-cooled bremsstrahlung radiator consisting of a 
$126$~{\textmu}m-thick Au foil and approximately $1$~cm of Cu backing. The 
resulting $2.52$~MeV endpoint bremsstrahlung photon beam was then collimated 
with
a 20~cm-long conical collimator of entry diameter $9.86$~mm and exit diameter 
$26.72$~mm, producing an opening half-angle of about~$5^\circ$. The beam 
configuration and stability are discussed in SI Appendix~\S 
\ref{sec:si_exp}.

Optional encryption plates directly after the collimator may be included as an 
additional layer of information security. The encryption plates are composed of 
warhead materials in amounts unknown to the inspector, so that any inference 
about the warhead composition will in fact be an inference on the warhead plus 
encryption plates, thus protecting the warhead information. As with the 
encryption foil, the encryption plates must remain constant between the template 
and candidate measurements. In these experiments, no such encryption plates were 
included in order to maximize the available flux and thus the statistical 
precision and sensitivity of the measurements.

After passing through the proxy warhead or hoax, the transmitted flux then 
impinged on the encryption foil, which was constructed from $3.18$~mm of DU 
plates followed by $63.5$~mm of aluminum plates. The uranium and aluminum 
components demonstrate the verification measurement for high- and low-$Z$ 
materials, respectively. Specifically, the measurements in this work are 
designed to show the
detection of high-$Z$ material diversions and the verification of low-$Z$ 
material consistency.

The combined NRF signature of the measurement target plus encryption foil---at 
this point physically encrypted---was measured using three mechanically cooled 
Ortec $100\%$ relative efficiency GEM P-type coaxial HPGe photon 
detectors. The detectors were placed ${\sim}45$~cm from the 
foil at an angle of 55$^\circ$ to the beam axis, and surrounded by lead to 
shield against NRF photons directly from the warhead, as well as active 
backgrounds from the experimental setting which would otherwise limit the 
performance of the detectors. The shielding moreover prevents the detectors 
from observing any passive photon spectra generated by radioactive material in 
the test objects. The lead shielding thickness ranged from $51$~mm below the 
detectors to $254$--$305$~mm along the line of sight from the collimator and 
warhead to the detectors. Only a $25.4$~mm lead filter was placed between the 
detectors and encryption foil. This reduced by multiple orders of magnitude the 
low energy photon flux, which can cause pileup and dead time in the detectors, 
with only a moderate reduction in the NRF signal. Finally, Canberra Lynx Digital 
Signal 
Analyzers were used to record the photon spectra in acquisition periods 
of five minutes (real time) in order to save the spectra for offline analysis 
and to estimate the detector dead time.

A $38.1$~mm right square cylinder lanthanum bromide (LaBr$_3$) 
crystal was placed downstream from the foil as an 
independent diagnostic for the bremsstrahlung beam flux. It should be 
emphasized 
that such additional measurements are not part of the verification protocol. 
They are, however, useful in an experimental setting for determining the 
bremsstrahlung endpoint energy of $2.52$~MeV (despite the $2.6$~MV reading of 
the accelerator terminal voltage---see SI Appendix \S \ref{sec:si_beam_char}) as small 
shifts in electron energy can have a large effect on absolute photon flux (and 
thus measurement time) near the 
endpoint. The LaBr$_3$ scintillator was chosen for its extremely fast decay time 
($16$~ns) and encased in a lead hut in order to avoid high pileup rates that 
could complicate the endpoint measurement---the detector was directly downbeam 
from the radiator, otherwise shielded only by the warhead and encryption foil. 
The detector was controlled using the ROOT-based~\cite{Brun1997} ADAQAcquisition 
software~\cite{hartwig2016adaq} and a CAEN DT-5790M digitizer.

\section*{Results and analysis}
For each measured object, photon spectra\footnote{Data and analysis code are available at https://github.com/jvavrek/PNAS2018} from multiple acquisition periods and 
three separate detectors are combined into a single 
live-charge-normalized\footnote{The term `live' is used to denote measurement 
times calculated using live time, i.e.,~the real time minus the detector's dead 
time. `Live charge' therefore corresponds to the product of beam current with 
live time.} spectrum in order to improve the signal-to-noise ratio (see SI Appendix \S 
\ref{sec:si_data}). Each spectrum is then fit with a series of Gaussian 
functions for the eight observed NRF peaks in the signal region near 
$2.1$--$2.3$~MeV, on top of an exponentially decaying continuum background. 
U-238 contributes the 2.176, 2.209, and 2.245~MeV peaks, the branched decays 
$45$~keV below each of these three, and a small peak with no branch at 
2.146~MeV. Al-27 contributes the intense 2.212~MeV peak. The Pb isotopes have no 
NRF lines below 2.3~MeV. Altogether, the spectral fitting function is written as
\begin{align}\label{eq:spectral_fit}
	\hspace*{-3pt}
    f(E) = \exp\left( c_1 + c_2 E \right) + \sum_{k=1}^8 
\frac{a_k}{\sqrt{2\pi}\sigma_k} \exp\left[ -\frac{(E-E_k)^2}{2\sigma_k^2} 
\right]
\end{align}
where $c_1$ and $c_2$ describe the shape of the continuum, and $a_k$, $E_k$, and 
$\sigma_k$ are the area, mean, and standard deviation fit parameters of the 
$k^\text{th}$ peak. With eight sets of three peak parameters and two parameters 
for the continuum, this results in a total of 26 parameters per spectrum.

Once the 26-parameter fit (and set of associated fit parameter uncertainties) 
for each spectrum is computed using Eq.~\ref{eq:spectral_fit}, the detected NRF 
rate in each peak in counts per live {\textmu}A$\cdot$s, as predicted by 
integration of SI Appendix Eq.~\ref{eq:d2ndEdOmega}, can be extracted as simply $A_k = a_k / 
\Delta E$, 
where $a_k$ is the value of the area fit parameter for the $k^\text{th}$ peak, 
and the division by the spectrum bin width $\Delta E$ enforces proper dimensions 
and normalization~\cite[p.~171]{bevington2003error}. Similarly, the uncertainty 
in the NRF rate is $\delta A_k = \delta a_k / \Delta E$ (where $\delta x$ is 
used to express the $1$~standard deviation 
uncertainty in a value $x$ so as to distinguish it from other uses of the 
symbol 
$\sigma$) where $\delta a_k$ is the uncertainty in the $a_k$ fit parameter as 
reported by ROOT's TH1::Fit() subroutines \cite{Brun1997}.

One possible test statistic $T$ for comparing the NRF 
peaks of a single isotope is the sum of net rates $A_k$ (above the fit 
background) of the six 
U-238 peaks well-separated from the doublet: $2.131$, $2.146$, $2.164$, 
$2.176$, 
$2.200$, and $2.245$~MeV.
The 2.209~MeV component of the 2.209 and 2.212 MeV doublet tends to have a 
larger uncertainty such that it does not contribute reliably to $T$, and thus 
is 
excluded. Moreover, since the amount of Al-27 (and the total high-$Z$ areal 
density) does not change between the warhead and hoax objects, the $2.212$~MeV 
peak rate is consistent throughout the measurements (up to day-to-day beam 
variations---see SI Appendix~\S\ref{sec:si_beam_char}). To compare the NRF spectrum of a 
candidate object to that of the genuine template, the discrepancy $\nu$ is 
defined as the difference in $T$ divided by the uncertainty in the 
difference:
\begin{align}\label{eq:discrep}
    \nu \equiv \frac{T_\text{cand}-T_\text{temp}}{\sqrt{(\delta 
T_\text{cand})^2 
+ (\delta T_\text{temp})^2}}.
\end{align}
As the presence of an NRF isotope in the object reduces the corresponding 
observed NRF rate (and thus $T$), $\nu > 0$ indicates a possible diversion of 
the isotope in the candidate compared to the template, while $\nu < 0$ indicates 
a possible addition. Under the null hypothesis that the 
candidate object is a real nuclear warhead, $T_\text{cand} = T_\text{temp}$, so 
that (due to statistics alone) $\nu$ is
normally distributed with mean 0 and standard deviation 1: $\nu \sim 
\mathcal{N}(0,1)$. As such, $\nu$ measures 
the discrepancy from the null hypothesis in number of  
standard deviations (``sigmas'') where, e.g.,~the probability of observing a 
$5\,\sigma$ discrepancy 
(regardless of sign) by chance alone, i.e.,~$|\nu| > 5$, is $6\times 10^{-7}$. 
Setting an alarm threshold $|\nu| > \nu^*$ by necessity trades-off
the probability that the measurement declares a genuine warhead to be a 
hoax (type~I error) and the probability that it declares a hoax warhead to be 
genuine (type~II error). If low type~I error is prioritized, a suitable alarm 
threshold may be $\nu^* = 5$, while $\nu^* = 3$ may be more suitable if low 
type~II error is desired.

\begin{figure}[t]
\centerline{\includegraphics[width=1.0\columnwidth]{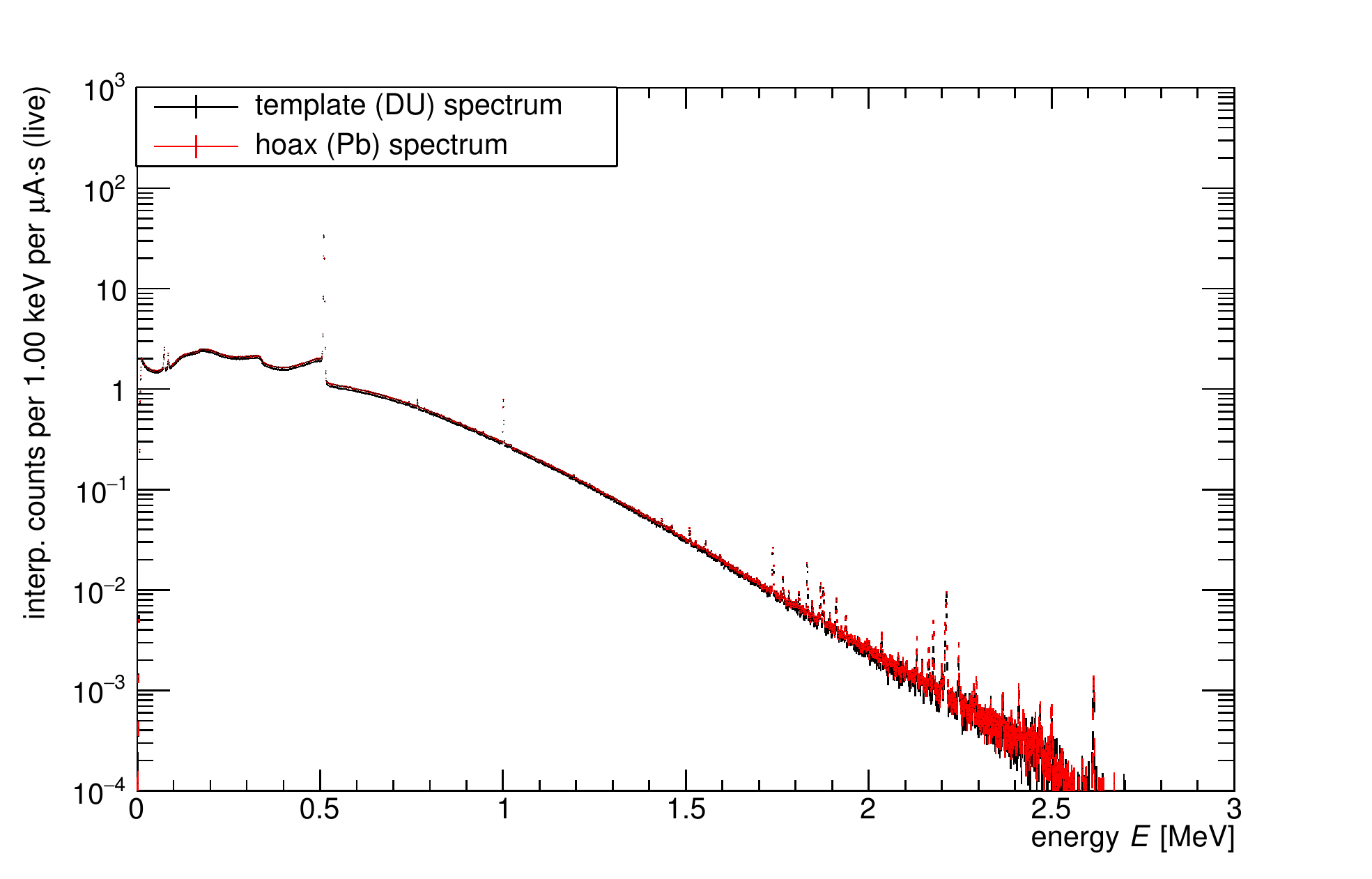}}
\caption{Measured spectra for DU template II (black points) and Pb hoax IIc 
(red 
points). In this and subsequent Figures, error bars are $\pm$1 SD.} 
\label{fig:c3_unzoom}
\end{figure}

\begin{figure}[t]
\centerline{\includegraphics[width=1.0\columnwidth]{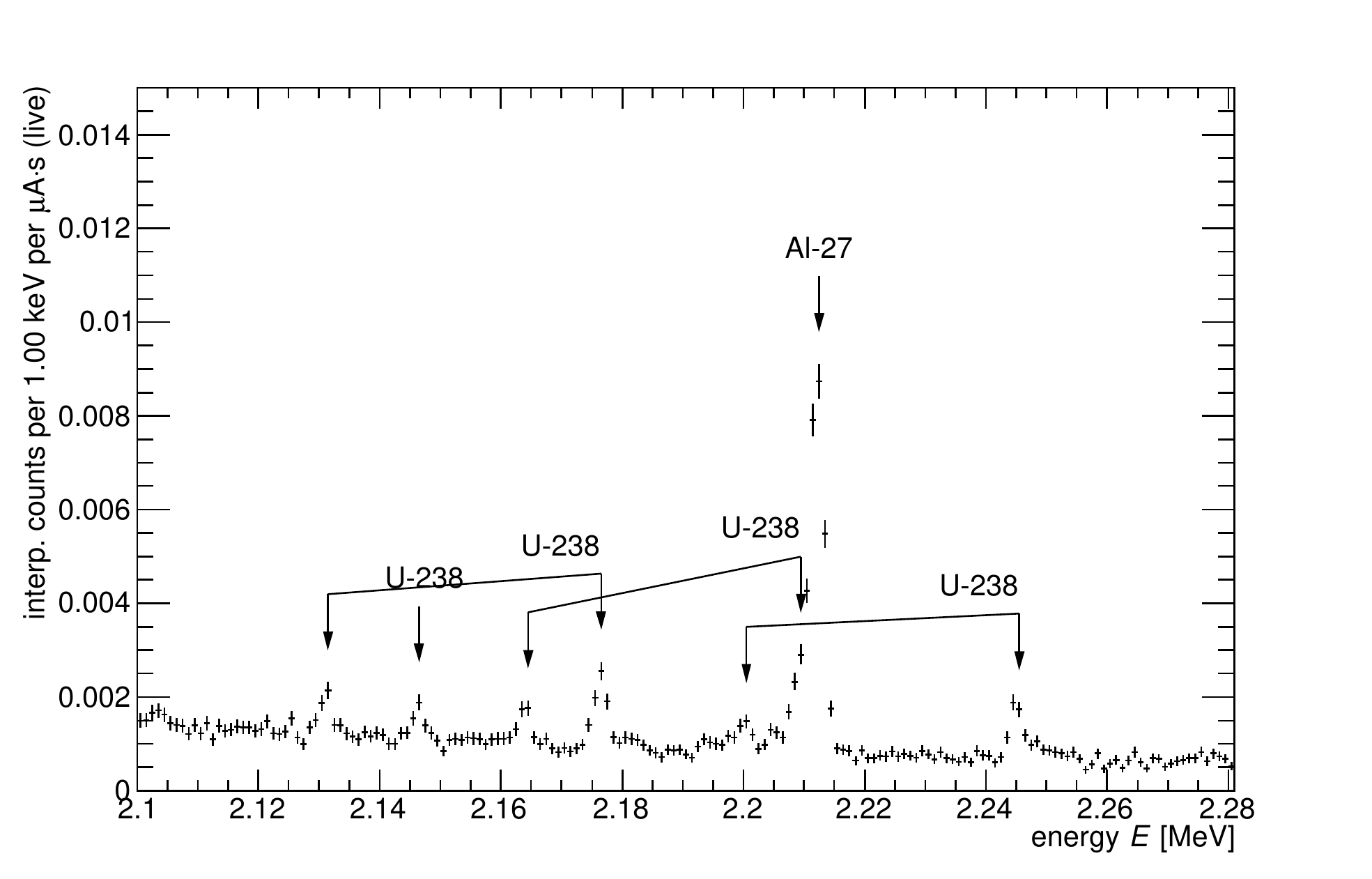}}
\caption{Measured spectra for DU template II, zoomed to show the NRF signal 
region. For clarity, the spectrum of hoax IIc is not shown. Arrows indicate the 
branching relationships from the three main U-238 lines to the peaks 45~keV 
lower, as well as the non-branching 2.146~MeV U-238 and 2.212~MeV Al-27 peaks.} 
\label{fig:c3_arrows}
\end{figure}

\begin{figure}[t]
\centerline{\includegraphics[width=1.0\columnwidth]{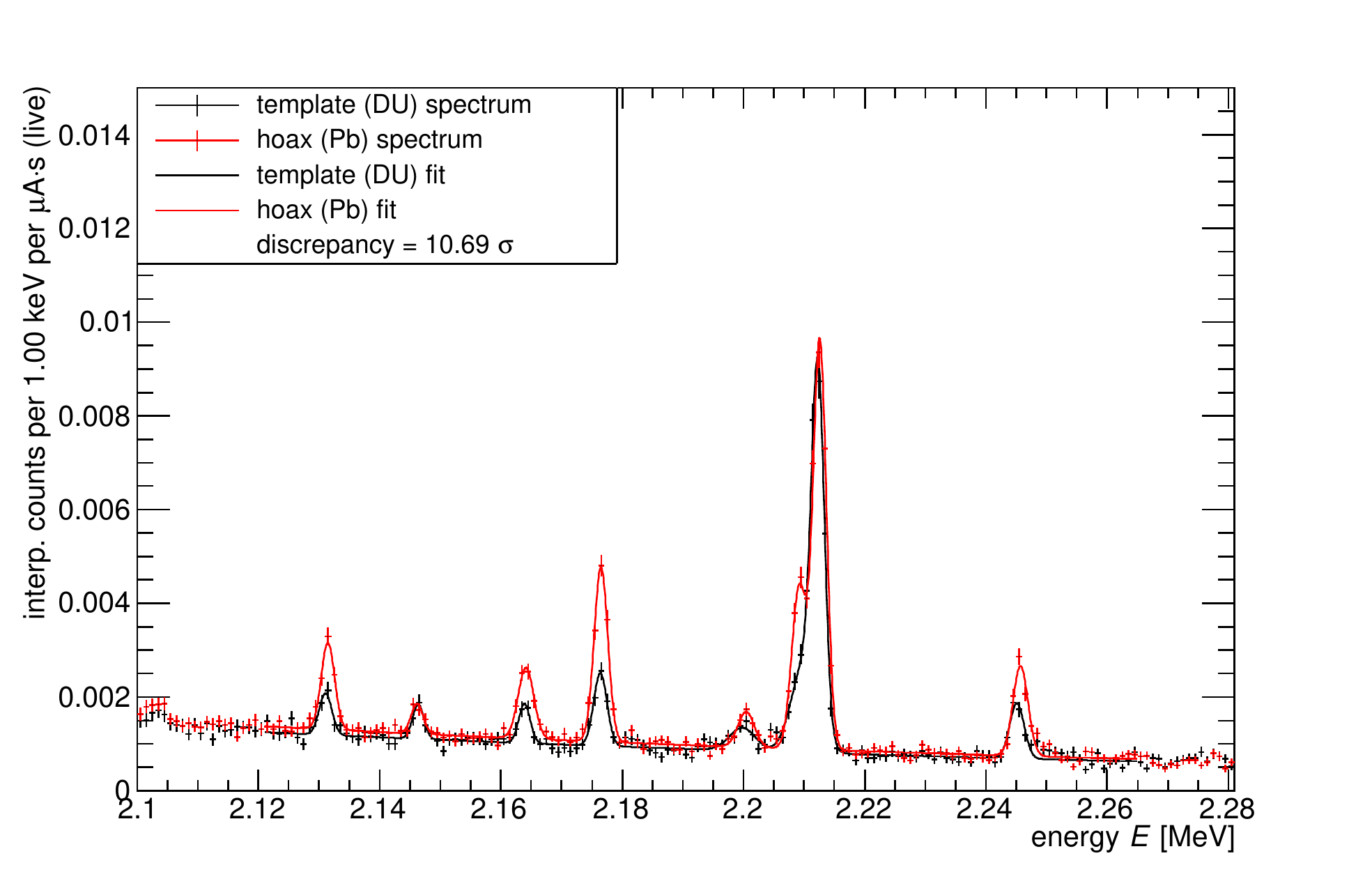}}
\caption{26-parameter Gaussian peak plus exponential background fits to the 
spectra of template II (black points and curve) and hoax IIc (red points and 
curve). A comparison of spectra for all verification measurements is shown in 
Table~\ref{tab:configs_small}.} \label{fig:c3_fit}
\end{figure}

Figs.~\ref{fig:c3_unzoom}, \ref{fig:c3_arrows}, and \ref{fig:c3_fit} show the
culmination of the above analysis procedure for the fourth verification scenario 
listed in Table~\ref{tab:configs_small} (template II vs hoax IIc). 
Fig.~\ref{fig:c3_unzoom} contains the two combined spectra measured for the 
template II (DU) and hoax IIc (Pb) proxy warheads; in this unzoomed energy 
range, the genuine and hoax spectra at first appear to match quite closely, 
with 
no obvious distinguishing features. Focusing on the NRF signal region in 
Fig.~\ref{fig:c3_arrows} (where only the template II spectrum is shown for 
clarity), the NRF peaks from U-238 and Al-27 become visible; 
Fig.~\ref{fig:c3_fit} subsequently shows the 26-parameter fits to the two 
spectra and the computed discrepancy of $\nu = 10.7$. The discrepancies for all 
verification scenarios are shown in Table~\ref{tab:configs_small} (see also 
SI Appendix Table~\ref{tab:configs_full}). In all four hoax scenarios, a discrepancy in $T$ 
greater than an alarm threshold of $\nu^* = 3$ was attained in 
${\sim}20$~{\textmu}A$\cdot$h (live, on three detectors) per
measured object, indicating diversions in the uranium component. In the genuine 
candidate scenario, the $1.7\, \sigma$ discrepancy in uranium (primarily a 
result of day-to-day beam variations) does not trigger the alarm at $\nu^* = 
3$, 
and is clearly delineated from the much larger observed discrepancies in the 
hoax cases. Similarly, the Al-27 comparisons all exhibit $|\nu| < 2$, 
indicating 
consistency in the aluminum component across all measurement scenarios.

The continua underlying the peaks---generated from both pileup and secondary 
electron bremsstrahlung in the foil---also provide some insight. For the 
spectra 
in Fig.~\ref{fig:c3_unzoom}, the integrals from $1$--$2$~MeV differ by 
$5\%$. The differences rise to $6$--$10\%$ when comparing measurements 
performed 
on different days due to beam variations, but are only $2\%$ in the other 
same-day measurements. Though these small differences are significant given the 
high statistics at low energies, the close matching of continua between the 
template and hoax scenarios suggests that the continuum background may not 
encode any appreciable information about the isotopic content of 
the weapon. This lack of distinguishing information in the 
majority of the spectrum also may indicate that non-resonant photon transmission
measurements such as radiographs would likely fail to detect hoaxes of the same areal density.

\begin{table}
\centering
\caption{Proxy warhead verification measurements}\label{tab:configs_small}
\begin{tabular}{cccc}
\hline\vspace{-2pt}
\# & scenario & Al-27 discrep. & U-238 discrep.\\
 - & - & $\nu$ (vs template) & $\nu$ (vs template)\\
\hline
0 & template I & - & -\\
1 & hoax Ia (100\% Pb) & $-$0.051~$\sigma$ & 7.9~$\sigma$\\
2 & genuine candidate Ig & 0.76~$\sigma$ & 1.7~$\sigma$\\
3 & hoax Ib (100\% Pb) & 1.7~$\sigma$ & 9.8~$\sigma$\\
4 & template II & - & -\\
5 & hoax IIc (100\% Pb) & 0.25~$\sigma$ & 10.7~$\sigma$\\
6 & hoax IId (50\% Pb) & 1.9~$\sigma$ & 4.6~$\sigma$\\
\hline
\end{tabular}
\end{table}

\section*{Discussion}
\subsection*{Extrapolation to future systems}
The proxy warheads used in this article are relatively thin---templates I and 
II 
have total on-axis areal densities of ${\sim}11$ and $17$~g/cm$^2$, 
respectively---and do not accurately represent typical areal densities of real 
warheads. More realistic warhead models in the open literature range from the 
compact 
(${\sim}50$~g/cm$^2$) Black Sea-type warhead used in~\cite{kemp2016physical} to 
the thicker (${\sim}200$~g/cm$^2$) models of 
Fetter \textit{et al.}~\cite{fetter1990detecting}. Moreover, verification 
measurements conducted on real warheads will use the NRF lines associated with 
the fuel isotopes U-235 or Pu-239, whose strongest lines are 
2--5$\times$ less intense than the U-238 lines 
considered in this work~\cite{ref:bertozzi_full}. In the present experimental 
design, 
verification of realistic weapon designs would therefore require several
orders of magnitude longer measurement times than the ${\sim}60$ detector live {\textmu}A$\cdot$h used here (see Table~\ref{tab:extrap_params}). In a dedicated warhead verification 
facility, however, these unrealistically long measurement times could be ameliorated by 
increasing the electron beam current and the number of detectors (see SI Appendix~\S\ref{sec:si_extrap}). A modern 
commercially available electron accelerator may have a continuous wave beam 
current of at least 5~mA at ${\sim}$3~MeV~\cite{dynamitron}, a factor of 
$200\times$ improvement over the 25~{\textmu}A used in this experiment. While
this increase in beam current would affect the event rate in the detectors (thus
reducing the effective live time), the increase in the event rate is only $\sim$30$\times$
due to the increased attenuation of realistic warheads (see SI Appendix~\S\ref{sec:si_extrap}).  This increase may be mitigated
by reducing the detector sizes and the operating with more detectors, by optimizing the balance of the
detector event rate and the available measurement time, or by taking advantage
of future developments of high-rate HPGe detectors capable of operating at MHz rates \cite{ref:pnnlhpge}.
Additionally, the shielding and detector filters used in this experiment could be signficantly
optimize to reduce the low energy photon rate in the HPGe detectors to further alleviate
this effect.
Extending the array of detectors from three to 30 would provide another factor 
of $10\times$ reduction in 
measurement time, and would provide the additional benefit of 
reducing the dose to the warhead---here estimated at ${\sim}30$~Gy per 1~hour 
measurement at 25~{\textmu}A for template~I---required to achieve the same 
confidence. Doses for other warhead configurations are presented in Ref.~\cite{kemp2016physical}. Such a dedicated 
verification system would be capable of 
attaining $5\,\sigma$ confidence in a single NRF line in a Pb 
hoax scenario involving the Fetter 
\textit{et al.} uranium-uranium model of Table~\ref{tab:extrap_params} in 
${\sim}15$--$20$~minutes per projection per object, for a capital cost on the order of USD~5M. 
This required runtime increases to ${\sim} 5$~hours for the worst-case 
plutonium-uranium model in Table~\ref{tab:extrap_params}. For thinner warheads, 
or for 
warheads that have been partially disassembled, even lower measurement times 
would be required, creating opportunities for measurements at multiple warhead 
orientations, for measurements of isotopes with weaker NRF lines, or for ruling 
out less discernible hoaxes.  More information on the calculation of the required
runtimes for realistic warhead configurations may be found in SI Appendix Section~\ref{sec:si_extrap}.

\begin{table}[hbt]
    \centering
    \caption{Warhead geometries and approximate detector live charges required 
for Pb replacement hoax detection at $5\,\sigma$.}\label{tab:extrap_params}
	\hspace*{-0.4cm}
	\scalebox{0.95}{
    \begin{tabular}{l|l|l|c}
comparison (model ref.) & NRF line & foil & det.~live {\textmu}A hr \\ 
\hline
        WGU+W  vs Pb+W~\cite{fetter1990detecting} & U-235 1.733 MeV & WGU & 
$25\times 10^3$\\
        WGU+DU vs Pb+DU~\cite{fetter1990detecting} & U-235 1.733 MeV & WGU & $40 
\times 10^3$\\
        WGPu+W  vs Pb+W~\cite{fetter1990detecting} & Pu-239 2.431 MeV & WGPu & 
$600 \times 10^3$\\
        WGPu+DU vs Pb+DU~\cite{fetter1990detecting} & Pu-239 2.431 MeV & WGPu & 
$800 \times 10^3$\\
        WGU  vs Pb~\cite{kemp2016physical} & U-235 1.733 MeV & WGU & $0.15 
\times 10^3$\\
        WGPu vs Pb~\cite{kemp2016physical} & Pu-239 2.431 MeV & WGPu & $3.5 
\times 10^3$
    \end{tabular}
    }
\end{table}

\subsection*{Information security}
The equation for the predicted NRF rates (SI Appendix Eq.~\ref{eq:d2ndEdOmega} or its 
integrated form), contains multiple quantities that are kept secret from the 
inspector, and thus cannot be used alone to infer the warhead thickness~$D$ 
from 
a physically encrypted spectrum. However, it may be possible to construct a 
system of equations from SI Appendix Eq.~\ref{eq:d2ndEdOmega}---one equation per NRF 
peak---and make a series of simplifying approximations, in which case there may 
be at least as many equations as unknowns and inference may be possible. As 
previously shown in Fig.~\ref{fig:schematic} and described in~\cite[SI~\S 
7.1]{kemp2016physical}, a solution to this \textit{multi-line inference} problem 
is to include optional encryption plates of relevant materials of unknown 
thickness $\Delta D$ to the collimator output. As such, any inference on the 
isotope of interest will estimate only an upper bound $D + \Delta D$. In fact, 
if such encryption plates are used, the foil parameter $X$ no longer needs to be 
kept secret from the inspector, eliminating the information security 
complexities of ensuring that the foil has not been nefariously designed.


Lastly, the continuum background may contain 
sensitive information, especially given the large number of photons it 
comprises 
over the entire range of the spectrum (see Fig.~\ref{fig:c3_unzoom}). The 
`logarithmic slope' parameter $c_2$ in Eq.~\ref{eq:spectral_fit}, for instance, 
depends moderately on the atomic number $Z$ of the foil 
materials~\cite{bertozzi2007ez3d}. As discussed above, however, the continuum 
appears to encode very little information about the $Z$ of the warhead materials 
for a fixed areal density.  A thorough analysis of the continuum information 
content is therefore a vital next step in the analysis of the physical 
cryptographic NRF protocol. For a more 
complete discussion of information security issues and possible solutions, the 
reader is referred to~\cite{kemp2016physical}.

\section*{Conclusions and future work}
We have reported on the successful demonstration of the MIT tNRF physical 
cryptographic warhead verification protocol. The isotope-sensitive tNRF 
measurement is capable of distinguishing proxy nuclear warheads from hoax 
objects with high confidence in total measurement times of around one hour per 
object. Extrapolations to more realistic warhead designs indicate that a 
dedicated warhead verification facility could conduct $5\,\sigma$ verification 
measurements in less than an hour while protecting sensitive warhead design 
information.

The NRF verification technique may be expanded to other isotopes 
that may be found in nuclear weapons (beyond U-238 and Al-27) such as U-235 or 
Pu-239 in the fissile fuel and nitrogen and carbon isotopes in the high 
explosives~\cite{caggiano2007nuclear}. Similarly, testing 
the measurement's sensitivity to geometric hoaxes would be a useful development.
Finally, an additional layer of information security may be added through 
analog-to-digital converters (ADCs) with non-uniform binning, which are 
currently being developed. Such ADCs would act as very low-level, more 
easily-verifiable information barriers. If installed in the acquisition systems of the HPGe detectors, such ADCs could be used to remove all spectral features 
except one NRF line per isotope from the observed spectrum, thus eliminating 
possible information security concerns such as the continuum and the multi-line 
inference problem.

In a broader context, the implementation of any warhead verification protocol in 
a real arms control agreement faces two challenges. First, an assessment of the 
protocol's utility and security must be made by nuclear weapons laboratories. To 
this end, future work on any warhead verification protocol should involve 
collaboration with the US and possibly Russian national laboratories, and 
possibly combining multiple proposed verification techniques as part of an 
overarching protocol. Such a joint effort will enable research that otherwise 
could not be conducted in academic settings, such as the aforementioned 
measurements involving weapons isotopes and realistic, classified weapon 
geometries. Finally, the implementation of a warhead verification protocol is 
predicated on the existence of future arms control frameworks, and thus requires 
a commitment to the goal of deep reductions in the world's nuclear arsenals.




%

\begin{acknowledgments}
 This work was funded in part by the Consortium for Verification Technology under Department of Energy National Nuclear Security Administration 
award number DE-NA0002534. BSH gratefully acknowledges the support of the 
Stanton Foundation's Nuclear Security Fellowship program. The authors thank 
Chathan Cooke for operating the HVRL accelerator, Charles Epstein and Richard 
Milner for their electron beam diagnostics, Rob Goldston for useful discussions, and R.~Scott Kemp for 
offering comments on the manuscript.
\end{acknowledgments}
\bibliography{bibliography}

\begin{thebibliography}{40}%
\makeatletter
\providecommand \@ifxundefined [1]{%
 \@ifx{#1\undefined}
}%
\providecommand \@ifnum [1]{%
 \ifnum #1\expandafter \@firstoftwo
 \else \expandafter \@secondoftwo
 \fi
}%
\providecommand \@ifx [1]{%
 \ifx #1\expandafter \@firstoftwo
 \else \expandafter \@secondoftwo
 \fi
}%
\providecommand \natexlab [1]{#1}%
\providecommand \enquote  [1]{``#1''}%
\providecommand \bibnamefont  [1]{#1}%
\providecommand \bibfnamefont [1]{#1}%
\providecommand \citenamefont [1]{#1}%
\providecommand \href@noop [0]{\@secondoftwo}%
\providecommand \href [0]{\begingroup \@sanitize@url \@href}%
\providecommand \@href[1]{\@@startlink{#1}\@@href}%
\providecommand \@@href[1]{\endgroup#1\@@endlink}%
\providecommand \@sanitize@url [0]{\catcode `\\12\catcode `\$12\catcode
  `\&12\catcode `\#12\catcode `\^12\catcode `\_12\catcode `\%12\relax}%
\providecommand \@@startlink[1]{}%
\providecommand \@@endlink[0]{}%
\providecommand \url  [0]{\begingroup\@sanitize@url \@url }%
\providecommand \@url [1]{\endgroup\@href {#1}{\urlprefix }}%
\providecommand \urlprefix  [0]{URL }%
\providecommand \Eprint [0]{\href }%
\providecommand \doibase [0]{http://dx.doi.org/}%
\providecommand \selectlanguage [0]{\@gobble}%
\providecommand \bibinfo  [0]{\@secondoftwo}%
\providecommand \bibfield  [0]{\@secondoftwo}%
\providecommand \translation [1]{[#1]}%
\providecommand \BibitemOpen [0]{}%
\providecommand \bibitemStop [0]{}%
\providecommand \bibitemNoStop [0]{.\EOS\space}%
\providecommand \EOS [0]{\spacefactor3000\relax}%
\providecommand \BibitemShut  [1]{\csname bibitem#1\endcsname}%
\let\auto@bib@innerbib\@empty
\bibitem [{\citenamefont {Drell}\ \emph {et~al.}(1993)\citenamefont {Drell},
  \citenamefont {Callan}, \citenamefont {Cornwall}, \citenamefont {Dyson},\
  and\ \citenamefont {Eardley}}]{drell1993verification}%
  \BibitemOpen
  \bibfield  {author} {\bibinfo {author} {\bibfnamefont {S.}~\bibnamefont
  {Drell}}, \bibinfo {author} {\bibfnamefont {C.}~\bibnamefont {Callan}},
  \bibinfo {author} {\bibfnamefont {J.}~\bibnamefont {Cornwall}}, \bibinfo
  {author} {\bibfnamefont {F.}~\bibnamefont {Dyson}}, \ and\ \bibinfo {author}
  {\bibfnamefont {D.}~\bibnamefont {Eardley}},\ }\href@noop {} {\emph {\bibinfo
  {title} {Verification of Dismantlement of Nuclear Warheads and Controls on
  Nuclear Materials}}},\ \bibinfo {type} {Tech. Rep.}\ \bibinfo {number}
  {JSR-92-331}\ (\bibinfo  {institution} {MITRE Corp JASON Program Office},\
  \bibinfo {address} {McLean, VA},\ \bibinfo {year} {1993})\BibitemShut
  {NoStop}%
\bibitem [{\citenamefont {Comley}\ \emph {et~al.}(2000)\citenamefont {Comley},
  \citenamefont {Comley},\ and\ \citenamefont {Eggins}}]{comley2000confidence}%
  \BibitemOpen
  \bibfield  {author} {\bibinfo {author} {\bibfnamefont {C.}~\bibnamefont
  {Comley}}, \bibinfo {author} {\bibfnamefont {M.}~\bibnamefont {Comley}}, \
  and\ \bibinfo {author} {\bibfnamefont {P.}~\bibnamefont {Eggins}},\
  }\href@noop {} {\emph {\bibinfo {title} {Confidence, security and
  verification: The challenge of global nuclear weapons arms control}}},\
  \bibinfo {type} {Tech. Rep.}\ \bibinfo {number} {AWE/TR/2000/001}\ (\bibinfo
  {institution} {United Kingdom Ministry of Defence},\ \bibinfo {address}
  {Whitehall, London},\ \bibinfo {year} {2000})\BibitemShut {NoStop}%
\bibitem [{\citenamefont {Spears}(2001)}]{spears2001technology}%
  \BibitemOpen
  \bibfield  {author} {\bibinfo {author} {\bibfnamefont {D.}~\bibnamefont
  {Spears}},\ }\href@noop {} {\bibfield  {journal} {\bibinfo  {journal} {US
  Department of Energy, National Nuclear Security Administration, Defense
  Nuclear Nonproliferation Programs, Washington, DC}\ } (\bibinfo {year}
  {2001})}\BibitemShut {NoStop}%
\bibitem [{\citenamefont {Fuller}(2010)}]{fuller}%
  \BibitemOpen
  \bibfield  {author} {\bibinfo {author} {\bibfnamefont {J.}~\bibnamefont
  {Fuller}},\ }\href@noop {} {\bibfield  {journal} {\bibinfo  {journal} {Arms
  Control Today}\ }\textbf {\bibinfo {volume} {40}},\ \bibinfo {pages} {19}
  (\bibinfo {year} {2010})}\BibitemShut {NoStop}%
\bibitem [{\citenamefont {Bunch}\ \emph {et~al.}(2014)\citenamefont {Bunch},
  \citenamefont {Jones}, \citenamefont {Ramuhalli}, \citenamefont {Benz},\ and\
  \citenamefont {Denlinger}}]{bunchCoC}%
  \BibitemOpen
  \bibfield  {author} {\bibinfo {author} {\bibfnamefont {K.~J.}\ \bibnamefont
  {Bunch}}, \bibinfo {author} {\bibfnamefont {M.}~\bibnamefont {Jones}},
  \bibinfo {author} {\bibfnamefont {P.}~\bibnamefont {Ramuhalli}}, \bibinfo
  {author} {\bibfnamefont {J.}~\bibnamefont {Benz}}, \ and\ \bibinfo {author}
  {\bibfnamefont {L.~S.}\ \bibnamefont {Denlinger}},\ }\href {\doibase
  10.1080/08929882.2014.917924} {\bibfield  {journal} {\bibinfo  {journal}
  {Sci.~Glob.~Sec.}\ }\textbf {\bibinfo {volume} {22}},\ \bibinfo {pages} {111}
  (\bibinfo {year} {2014})},\ \Eprint
  {http://arxiv.org/abs/https://doi.org/10.1080/08929882.2014.917924}
  {https://doi.org/10.1080/08929882.2014.917924} \BibitemShut {NoStop}%
\bibitem [{\citenamefont {{Defense Threat Reduction
  Agency}}(2001)}]{lanl2001fmttd}%
  \BibitemOpen
  \bibfield  {author} {\bibinfo {author} {\bibnamefont {{Defense Threat
  Reduction Agency}}},\ }\href@noop {} {\emph {\bibinfo {title} {Technical
  Overview of {Fissile Material Transparency Technology Demonstration}}}},\
  \bibinfo {type} {Tech. Rep.}\ (\bibinfo  {institution} {Los Alamos National
  Laboratory},\ \bibinfo {address} {Los Alamos, NM},\ \bibinfo {year} {2001})\
  \bibinfo {note} {retrieved from
  \url{http://www.lanl.gov/orgs/n/n1/FMTTD/presentations/pdf\_docs/exec\_sum.pdf}
  on December 4, 2017.}\BibitemShut {Stop}%
\bibitem [{\citenamefont {Close}\ \emph {et~al.}(2001)\citenamefont {Close},
  \citenamefont {MacArthur},\ and\ \citenamefont
  {Nicholas}}]{close2001infobarriers}%
  \BibitemOpen
  \bibfield  {author} {\bibinfo {author} {\bibfnamefont {D.}~\bibnamefont
  {Close}}, \bibinfo {author} {\bibfnamefont {D.}~\bibnamefont {MacArthur}}, \
  and\ \bibinfo {author} {\bibfnamefont {N.}~\bibnamefont {Nicholas}},\
  }\href@noop {} {\emph {\bibinfo {title} {Information Barriers---A Historical
  Perspective}}},\ \bibinfo {type} {Tech. Rep.}\ \bibinfo {number}
  {LA-UR-01-2180}\ (\bibinfo  {institution} {Los Alamos National Laboratory},\
  \bibinfo {address} {Los Alamos, NM},\ \bibinfo {year} {2001})\ \bibinfo
  {note} {retrieved from \url{http://lib-www.lanl.gov/la-pubs/00796106.pdf} on
  December 4, 2017.}\BibitemShut {Stop}%
\bibitem [{\citenamefont {Yan}\ and\ \citenamefont
  {Glaser}(2015{\natexlab{a}})}]{yan2015review}%
  \BibitemOpen
  \bibfield  {author} {\bibinfo {author} {\bibfnamefont {J.}~\bibnamefont
  {Yan}}\ and\ \bibinfo {author} {\bibfnamefont {A.}~\bibnamefont {Glaser}},\
  }\href
  {http://scienceandglobalsecurity.org/archive/2015/09/nuclear_warhead_verification_a.html}
  {\bibfield  {journal} {\bibinfo  {journal} {Sci.~Glob.~Sec.}\ }\textbf
  {\bibinfo {volume} {23}},\ \bibinfo {pages} {157} (\bibinfo {year}
  {2015}{\natexlab{a}})},\ \Eprint
  {http://arxiv.org/abs/http://scienceandglobalsecurity.org/archive/sgs23jieyan.pdf}
  {http://scienceandglobalsecurity.org/archive/sgs23jieyan.pdf} \BibitemShut
  {NoStop}%
\bibitem [{\citenamefont {Marleau}\ \emph {et~al.}(2015)\citenamefont
  {Marleau}, \citenamefont {Brubaker}, \citenamefont {Deland}, \citenamefont
  {Hilton}, \citenamefont {McDaniel}, \citenamefont {Schroeppel}, \citenamefont
  {Seager}, \citenamefont {Stoddard},\ and\ \citenamefont
  {MacArthur}}]{marleau2015tabletop}%
  \BibitemOpen
  \bibfield  {author} {\bibinfo {author} {\bibfnamefont {P.}~\bibnamefont
  {Marleau}}, \bibinfo {author} {\bibfnamefont {E.}~\bibnamefont {Brubaker}},
  \bibinfo {author} {\bibfnamefont {S.}~\bibnamefont {Deland}}, \bibinfo
  {author} {\bibfnamefont {N.}~\bibnamefont {Hilton}}, \bibinfo {author}
  {\bibfnamefont {M.}~\bibnamefont {McDaniel}}, \bibinfo {author}
  {\bibfnamefont {R.}~\bibnamefont {Schroeppel}}, \bibinfo {author}
  {\bibfnamefont {K.}~\bibnamefont {Seager}}, \bibinfo {author} {\bibfnamefont
  {M.~C.}\ \bibnamefont {Stoddard}}, \ and\ \bibinfo {author} {\bibfnamefont
  {D.}~\bibnamefont {MacArthur}},\ }\href@noop {} {\emph {\bibinfo {title}
  {Report on a zero-knowledge protocol tabletop exercise}}},\ \bibinfo {type}
  {Tech. Rep.}\ \bibinfo {number} {SAND2015-5075}\ (\bibinfo  {institution}
  {Sandia National Laboratories, Los Alamos National Laboratories},\ \bibinfo
  {address} {Livermore, CA and Los Alamos, NM},\ \bibinfo {year}
  {2015})\BibitemShut {NoStop}%
\bibitem [{\citenamefont {Glaser}\ \emph {et~al.}(2014)\citenamefont {Glaser},
  \citenamefont {Barak},\ and\ \citenamefont {Goldston}}]{ref:alex}%
  \BibitemOpen
  \bibfield  {author} {\bibinfo {author} {\bibfnamefont {A.}~\bibnamefont
  {Glaser}}, \bibinfo {author} {\bibfnamefont {B.}~\bibnamefont {Barak}}, \
  and\ \bibinfo {author} {\bibfnamefont {R.}~\bibnamefont {Goldston}},\
  }\href@noop {} {\bibfield  {journal} {\bibinfo  {journal} {Nature}\ }\textbf
  {\bibinfo {volume} {510}},\ \bibinfo {pages} {497} (\bibinfo {year}
  {2014})}\BibitemShut {NoStop}%
\bibitem [{\citenamefont {Philippe}\ \emph {et~al.}(2016)\citenamefont
  {Philippe}, \citenamefont {Goldston}, \citenamefont {Glaser},\ and\
  \citenamefont {d’Errico}}]{philippe2016verification}%
  \BibitemOpen
  \bibfield  {author} {\bibinfo {author} {\bibfnamefont {S.}~\bibnamefont
  {Philippe}}, \bibinfo {author} {\bibfnamefont {R.~J.}\ \bibnamefont
  {Goldston}}, \bibinfo {author} {\bibfnamefont {A.}~\bibnamefont {Glaser}}, \
  and\ \bibinfo {author} {\bibfnamefont {F.}~\bibnamefont {d’Errico}},\
  }\href@noop {} {\bibfield  {journal} {\bibinfo  {journal} {Nat Commun}\
  }\textbf {\bibinfo {volume} {7}},\ \bibinfo {pages} {12890} (\bibinfo {year}
  {2016})}\BibitemShut {NoStop}%
\bibitem [{\citenamefont {Kemp}\ \emph {et~al.}(2016)\citenamefont {Kemp},
  \citenamefont {Danagoulian}, \citenamefont {Macdonald},\ and\ \citenamefont
  {Vavrek}}]{kemp2016physical}%
  \BibitemOpen
  \bibfield  {author} {\bibinfo {author} {\bibfnamefont {R.~S.}\ \bibnamefont
  {Kemp}}, \bibinfo {author} {\bibfnamefont {A.}~\bibnamefont {Danagoulian}},
  \bibinfo {author} {\bibfnamefont {R.~R.}\ \bibnamefont {Macdonald}}, \ and\
  \bibinfo {author} {\bibfnamefont {J.~R.}\ \bibnamefont {Vavrek}},\
  }\href@noop {} {\bibfield  {journal} {\bibinfo  {journal} {Proc Natl Acad Sci
  USA}\ }\textbf {\bibinfo {volume} {113}},\ \bibinfo {pages} {8618} (\bibinfo
  {year} {2016})}\BibitemShut {NoStop}%
\bibitem [{\citenamefont {Marleau}\ and\ \citenamefont
  {Krentz-Wee}(2017)}]{marleau2017implementations}%
  \BibitemOpen
  \bibfield  {author} {\bibinfo {author} {\bibfnamefont {P.}~\bibnamefont
  {Marleau}}\ and\ \bibinfo {author} {\bibfnamefont {R.}~\bibnamefont
  {Krentz-Wee}},\ }\href@noop {} {\emph {\bibinfo {title} {Investigation into
  practical implementations of a zero knowledge protocol}}},\ \bibinfo {type}
  {Tech. Rep.}\ \bibinfo {number} {SAND2017-1649}\ (\bibinfo  {institution}
  {Sandia National Laboratories},\ \bibinfo {address} {Livermore, CA},\
  \bibinfo {year} {2017})\BibitemShut {NoStop}%
\bibitem [{\citenamefont {Hecla}\ and\ \citenamefont
  {Danagoulian}(2017)}]{hecla2017epithermal}%
  \BibitemOpen
  \bibfield  {author} {\bibinfo {author} {\bibfnamefont {J.~J.}\ \bibnamefont
  {Hecla}}\ and\ \bibinfo {author} {\bibfnamefont {A.}~\bibnamefont
  {Danagoulian}},\ }\href@noop {} {\bibfield  {journal} {\bibinfo  {journal}
  {arXiv preprint arXiv:1709.09736}\ } (\bibinfo {year} {2017})}\BibitemShut
  {NoStop}%
\bibitem [{\citenamefont {Goldwasser}\ \emph {et~al.}(1989)\citenamefont
  {Goldwasser}, \citenamefont {Micali},\ and\ \citenamefont
  {Rackoff}}]{goldwasser1989knowledge}%
  \BibitemOpen
  \bibfield  {author} {\bibinfo {author} {\bibfnamefont {S.}~\bibnamefont
  {Goldwasser}}, \bibinfo {author} {\bibfnamefont {S.}~\bibnamefont {Micali}},
  \ and\ \bibinfo {author} {\bibfnamefont {C.}~\bibnamefont {Rackoff}},\
  }\href@noop {} {\bibfield  {journal} {\bibinfo  {journal} {SIAM Journal on
  Computing}\ }\textbf {\bibinfo {volume} {18}},\ \bibinfo {pages} {186}
  (\bibinfo {year} {1989})}\BibitemShut {NoStop}%
\bibitem [{\citenamefont {Blum}\ \emph {et~al.}(1988)\citenamefont {Blum},
  \citenamefont {Feldman},\ and\ \citenamefont {Micali}}]{blum1988zk}%
  \BibitemOpen
  \bibfield  {author} {\bibinfo {author} {\bibfnamefont {M.}~\bibnamefont
  {Blum}}, \bibinfo {author} {\bibfnamefont {P.}~\bibnamefont {Feldman}}, \
  and\ \bibinfo {author} {\bibfnamefont {S.}~\bibnamefont {Micali}},\ }in\
  \href@noop {} {\emph {\bibinfo {booktitle} {Proc.~of the
  20\textsuperscript{th} Annual Association for Computing Machinery Symposium
  on Theory of Computing}}}\ (\bibinfo {address} {Chicago, IL},\ \bibinfo
  {year} {1988})\ pp.\ \bibinfo {pages} {103--112}\BibitemShut {NoStop}%
\bibitem [{\citenamefont {Yan}\ and\ \citenamefont
  {Glaser}(2015{\natexlab{b}})}]{yan2015two}%
  \BibitemOpen
  \bibfield  {author} {\bibinfo {author} {\bibfnamefont {J.}~\bibnamefont
  {Yan}}\ and\ \bibinfo {author} {\bibfnamefont {A.}~\bibnamefont {Glaser}},\
  }in\ \href@noop {} {\emph {\bibinfo {booktitle} {Proc. 56\textsuperscript{th}
  Annual INMM Meeting}}}\ (\bibinfo {year} {2015})\ \bibinfo {note} {retrieved
  from \url{http://www.princeton.edu/~aglaser/PU105-Yan-Jie-Glaser-2015.pdf} on
  Oct.~23, 2017.}\BibitemShut {Stop}%
\bibitem [{\citenamefont {Metzger}(1959)}]{metzger1959resonance}%
  \BibitemOpen
  \bibfield  {author} {\bibinfo {author} {\bibfnamefont {F.}~\bibnamefont
  {Metzger}},\ }\href@noop {} {\bibfield  {journal} {\bibinfo  {journal} {Prog
  Nucl Phys}\ }\textbf {\bibinfo {volume} {7}},\ \bibinfo {pages} {54}
  (\bibinfo {year} {1959})}\BibitemShut {NoStop}%
\bibitem [{\citenamefont {Kneissl}\ \emph {et~al.}(1996)\citenamefont
  {Kneissl}, \citenamefont {Pitz},\ and\ \citenamefont
  {Zilges}}]{kneissl1996structure}%
  \BibitemOpen
  \bibfield  {author} {\bibinfo {author} {\bibfnamefont {U.}~\bibnamefont
  {Kneissl}}, \bibinfo {author} {\bibfnamefont {H.}~\bibnamefont {Pitz}}, \
  and\ \bibinfo {author} {\bibfnamefont {A.}~\bibnamefont {Zilges}},\
  }\href@noop {} {\bibfield  {journal} {\bibinfo  {journal} {Prog Part Nucl
  Phys}\ }\textbf {\bibinfo {volume} {37}},\ \bibinfo {pages} {349} (\bibinfo
  {year} {1996})}\BibitemShut {NoStop}%
\bibitem [{\citenamefont {Brun}\ and\ \citenamefont
  {Rademakers}(1997)}]{Brun1997}%
  \BibitemOpen
  \bibfield  {author} {\bibinfo {author} {\bibfnamefont {R.}~\bibnamefont
  {Brun}}\ and\ \bibinfo {author} {\bibfnamefont {F.}~\bibnamefont
  {Rademakers}} (\bibinfo {collaboration} {ROOT}),\ }\href@noop {} {\bibfield
  {journal} {\bibinfo  {journal} {Nucl Instrum Methods in Phys Res}\ }\textbf
  {\bibinfo {volume} {A 389}},\ \bibinfo {pages} {81} (\bibinfo {year}
  {1997})}\BibitemShut {NoStop}%
\bibitem [{\citenamefont {Hartwig}(2016)}]{hartwig2016adaq}%
  \BibitemOpen
  \bibfield  {author} {\bibinfo {author} {\bibfnamefont {Z.~S.}\ \bibnamefont
  {Hartwig}},\ }\href@noop {} {\bibfield  {journal} {\bibinfo  {journal} {Nucl
  Instrum Methods Phys Res A}\ }\textbf {\bibinfo {volume} {815}},\ \bibinfo
  {pages} {42} (\bibinfo {year} {2016})}\BibitemShut {NoStop}%
\bibitem [{\citenamefont {Bevington}\ and\ \citenamefont
  {Robinson}(2003)}]{bevington2003error}%
  \BibitemOpen
  \bibfield  {author} {\bibinfo {author} {\bibfnamefont {P.~R.}\ \bibnamefont
  {Bevington}}\ and\ \bibinfo {author} {\bibfnamefont {D.~K.}\ \bibnamefont
  {Robinson}},\ }\href@noop {} {\emph {\bibinfo {title} {Data reduction and
  error analysis for the physical sciences}}}\ (\bibinfo  {publisher}
  {McGraw-Hill},\ \bibinfo {address} {New York, NY},\ \bibinfo {year}
  {2003})\BibitemShut {NoStop}%
\bibitem [{\citenamefont {Fetter}\ \emph
  {et~al.}(1990{\natexlab{a}})\citenamefont {Fetter}, \citenamefont {Frolov},
  \citenamefont {Miller}, \citenamefont {Mozley}, \citenamefont {Prilutsky},
  \citenamefont {Rodionov},\ and\ \citenamefont
  {Sagdeev}}]{fetter1990detecting}%
  \BibitemOpen
  \bibfield  {author} {\bibinfo {author} {\bibfnamefont {S.}~\bibnamefont
  {Fetter}}, \bibinfo {author} {\bibfnamefont {V.~A.}\ \bibnamefont {Frolov}},
  \bibinfo {author} {\bibfnamefont {M.}~\bibnamefont {Miller}}, \bibinfo
  {author} {\bibfnamefont {R.}~\bibnamefont {Mozley}}, \bibinfo {author}
  {\bibfnamefont {O.~F.}\ \bibnamefont {Prilutsky}}, \bibinfo {author}
  {\bibfnamefont {S.~N.}\ \bibnamefont {Rodionov}}, \ and\ \bibinfo {author}
  {\bibfnamefont {R.~Z.}\ \bibnamefont {Sagdeev}},\ }\href@noop {} {\bibfield
  {journal} {\bibinfo  {journal} {Sci.~Glob.~Sec.}\ }\textbf {\bibinfo {volume}
  {1}},\ \bibinfo {pages} {225} (\bibinfo {year}
  {1990}{\natexlab{a}})}\BibitemShut {NoStop}%
\bibitem [{\citenamefont {Bertozzi}\ \emph {et~al.}(2008)\citenamefont
  {Bertozzi}, \citenamefont {Caggiano}, \citenamefont {Hensley}, \citenamefont
  {Johnson}, \citenamefont {Korbly}, \citenamefont {Ledoux}, \citenamefont
  {McNabb}, \citenamefont {Norman}, \citenamefont {Park},\ and\ \citenamefont
  {Warren}}]{ref:bertozzi_full}%
  \BibitemOpen
  \bibfield  {author} {\bibinfo {author} {\bibfnamefont {W.}~\bibnamefont
  {Bertozzi}}, \bibinfo {author} {\bibfnamefont {J.}~\bibnamefont {Caggiano}},
  \bibinfo {author} {\bibfnamefont {W.}~\bibnamefont {Hensley}}, \bibinfo
  {author} {\bibfnamefont {M.}~\bibnamefont {Johnson}}, \bibinfo {author}
  {\bibfnamefont {S.}~\bibnamefont {Korbly}}, \bibinfo {author} {\bibfnamefont
  {R.}~\bibnamefont {Ledoux}}, \bibinfo {author} {\bibfnamefont
  {D.}~\bibnamefont {McNabb}}, \bibinfo {author} {\bibfnamefont
  {E.}~\bibnamefont {Norman}}, \bibinfo {author} {\bibfnamefont
  {W.}~\bibnamefont {Park}}, \ and\ \bibinfo {author} {\bibfnamefont
  {G.}~\bibnamefont {Warren}},\ }\href {\doibase 10.1103/PhysRevC.78.041601}
  {\bibfield  {journal} {\bibinfo  {journal} {Phys Rev C}\ }\textbf {\bibinfo
  {volume} {78}},\ \bibinfo {pages} {041601} (\bibinfo {year}
  {2008})}\BibitemShut {NoStop}%
\bibitem [{\citenamefont {{IBA Industrial}}(2017)}]{dynamitron}%
  \BibitemOpen
  \bibfield  {author} {\bibinfo {author} {\bibnamefont {{IBA Industrial}}},\
  }\href@noop {} {\enquote {\bibinfo {title} {{IBA Dynamitron}},}\ } (\bibinfo
  {year} {2017}),\ \bibinfo {note} {retrieved from
  \url{http://www.iba-industrial.com/accelerators\#dynamitron-e-beam-accelerator}
  on Oct.~20, 2017}\BibitemShut {NoStop}%
\bibitem [{\citenamefont {Fast}\ \emph {et~al.}(2013)\citenamefont {Fast},
  \citenamefont {Dion}, \citenamefont {Rodriguez}, \citenamefont {VanDevender},
  \citenamefont {Wood},\ and\ \citenamefont {Wright}}]{ref:pnnlhpge}%
  \BibitemOpen
  \bibfield  {author} {\bibinfo {author} {\bibfnamefont {J.}~\bibnamefont
  {Fast}}, \bibinfo {author} {\bibfnamefont {M.}~\bibnamefont {Dion}}, \bibinfo
  {author} {\bibfnamefont {D.}~\bibnamefont {Rodriguez}}, \bibinfo {author}
  {\bibfnamefont {B.}~\bibnamefont {VanDevender}}, \bibinfo {author}
  {\bibfnamefont {L.}~\bibnamefont {Wood}}, \ and\ \bibinfo {author}
  {\bibfnamefont {M.}~\bibnamefont {Wright}},\ }\href@noop {} {\emph {\bibinfo
  {title} {{Performance of the Ultra-High Rate Germanium (UHRGe) System}}}},\
  \bibinfo {type} {Tech. Rep.}\ \bibinfo {number} {PNNL-23084}\ (\bibinfo
  {institution} {Pacific Northwest National Laboratory},\ \bibinfo {address}
  {Richland, WA},\ \bibinfo {year} {2013})\BibitemShut {NoStop}%
\bibitem [{\citenamefont {Bertozzi}\ \emph {et~al.}(2007)\citenamefont
  {Bertozzi}, \citenamefont {Korbly}, \citenamefont {Ledoux},\ and\
  \citenamefont {Park}}]{bertozzi2007ez3d}%
  \BibitemOpen
  \bibfield  {author} {\bibinfo {author} {\bibfnamefont {W.}~\bibnamefont
  {Bertozzi}}, \bibinfo {author} {\bibfnamefont {S.~E.}\ \bibnamefont
  {Korbly}}, \bibinfo {author} {\bibfnamefont {R.~J.}\ \bibnamefont {Ledoux}},
  \ and\ \bibinfo {author} {\bibfnamefont {W.}~\bibnamefont {Park}},\
  }\href@noop {} {\bibfield  {journal} {\bibinfo  {journal} {Nucl Instrum
  Methods Phys Res B}\ }\textbf {\bibinfo {volume} {261}},\ \bibinfo {pages}
  {331} (\bibinfo {year} {2007})}\BibitemShut {NoStop}%
\bibitem [{\citenamefont {Caggiano}\ \emph {et~al.}(2007)\citenamefont
  {Caggiano}, \citenamefont {Warren}, \citenamefont {Korbly}, \citenamefont
  {Hasty}, \citenamefont {Klimenko},\ and\ \citenamefont
  {Park}}]{caggiano2007nuclear}%
  \BibitemOpen
  \bibfield  {author} {\bibinfo {author} {\bibfnamefont {J.~A.}\ \bibnamefont
  {Caggiano}}, \bibinfo {author} {\bibfnamefont {G.~A.}\ \bibnamefont
  {Warren}}, \bibinfo {author} {\bibfnamefont {S.~E.}\ \bibnamefont {Korbly}},
  \bibinfo {author} {\bibfnamefont {R.~A.}\ \bibnamefont {Hasty}}, \bibinfo
  {author} {\bibfnamefont {A.}~\bibnamefont {Klimenko}}, \ and\ \bibinfo
  {author} {\bibfnamefont {W.~H.}\ \bibnamefont {Park}},\ }in\ \href@noop {}
  {\emph {\bibinfo {booktitle} {Nuclear Science Symposium Conference Record,
  2007. NSS'07. IEEE}}},\ Vol.~\bibinfo {volume} {3}\ (\bibinfo {organization}
  {IEEE},\ \bibinfo {address} {Honolulu, HI},\ \bibinfo {year} {2007})\ pp.\
  \bibinfo {pages} {2045--2046}\BibitemShut {NoStop}%
\bibitem [{\citenamefont {Browne}\ and\ \citenamefont
  {Tuli}(2015)}]{nndc2015u238}%
  \BibitemOpen
  \bibfield  {author} {\bibinfo {author} {\bibfnamefont {E.}~\bibnamefont
  {Browne}}\ and\ \bibinfo {author} {\bibfnamefont {J.}~\bibnamefont {Tuli}},\
  }\href@noop {} {\bibfield  {journal} {\bibinfo  {journal} {Nuclear Data
  Sheets}\ }\textbf {\bibinfo {volume} {127}},\ \bibinfo {pages} {191}
  (\bibinfo {year} {2015})}\BibitemShut {NoStop}%
\bibitem [{\citenamefont {Pruet}\ \emph {et~al.}(2006)\citenamefont {Pruet},
  \citenamefont {McNabb}, \citenamefont {Hagmann}, \citenamefont {Hartemann},\
  and\ \citenamefont {Barty}}]{pruet2006detecting}%
  \BibitemOpen
  \bibfield  {author} {\bibinfo {author} {\bibfnamefont {J.}~\bibnamefont
  {Pruet}}, \bibinfo {author} {\bibfnamefont {D.}~\bibnamefont {McNabb}},
  \bibinfo {author} {\bibfnamefont {C.}~\bibnamefont {Hagmann}}, \bibinfo
  {author} {\bibfnamefont {F.}~\bibnamefont {Hartemann}}, \ and\ \bibinfo
  {author} {\bibfnamefont {C.}~\bibnamefont {Barty}},\ }\href@noop {}
  {\bibfield  {journal} {\bibinfo  {journal} {Journal of Applied Physics}\
  }\textbf {\bibinfo {volume} {99}},\ \bibinfo {pages} {123102} (\bibinfo
  {year} {2006})}\BibitemShut {NoStop}%
\bibitem [{\citenamefont {Quiter}\ \emph {et~al.}(2011)\citenamefont {Quiter},
  \citenamefont {Ludewigt}, \citenamefont {Mozin}, \citenamefont {Wilson},\
  and\ \citenamefont {Korbly}}]{ref:quiter}%
  \BibitemOpen
  \bibfield  {author} {\bibinfo {author} {\bibfnamefont {B.~J.}\ \bibnamefont
  {Quiter}}, \bibinfo {author} {\bibfnamefont {B.~A.}\ \bibnamefont
  {Ludewigt}}, \bibinfo {author} {\bibfnamefont {V.~V.}\ \bibnamefont {Mozin}},
  \bibinfo {author} {\bibfnamefont {C.}~\bibnamefont {Wilson}}, \ and\ \bibinfo
  {author} {\bibfnamefont {S.}~\bibnamefont {Korbly}},\ }\href@noop {}
  {\bibfield  {journal} {\bibinfo  {journal} {Nucl Instrum Methods Phys Res B}\
  }\textbf {\bibinfo {volume} {269}},\ \bibinfo {pages} {1130 } (\bibinfo
  {year} {2011})}\BibitemShut {NoStop}%
\bibitem [{\citenamefont {Hamilton}(1940)}]{hamilton1940directional}%
  \BibitemOpen
  \bibfield  {author} {\bibinfo {author} {\bibfnamefont {D.~R.}\ \bibnamefont
  {Hamilton}},\ }\href@noop {} {\bibfield  {journal} {\bibinfo  {journal}
  {Physical Review}\ }\textbf {\bibinfo {volume} {58}},\ \bibinfo {pages} {122}
  (\bibinfo {year} {1940})}\BibitemShut {NoStop}%
\bibitem [{\citenamefont {Quiter}(2010)}]{quiter2010thesis}%
  \BibitemOpen
  \bibfield  {author} {\bibinfo {author} {\bibfnamefont {B.}~\bibnamefont
  {Quiter}},\ }\emph {\bibinfo {title} {Nuclear resonance fluorescence for
  nuclear materials assay}},\ \href@noop {} {Ph.D. thesis},\ \bibinfo  {school}
  {University of California, Berkeley} (\bibinfo {year} {2010})\BibitemShut
  {NoStop}%
\bibitem [{\citenamefont {{AMETEK Inc.}}(2017)}]{ortec_gem}%
  \BibitemOpen
  \bibfield  {author} {\bibinfo {author} {\bibnamefont {{AMETEK Inc.}}},\
  }\href@noop {} {\enquote {\bibinfo {title} {{GEM} series coaxial {HPGe}
  detector product configuration guide},}\ } (\bibinfo {year} {2017}),\
  \bibinfo {note} {retrieved from
  \url{http://www.ortec-online.com/-/media/ametekortec/brochures/gem.pdf} on
  Oct.~23, 2017}\BibitemShut {NoStop}%
\bibitem [{\citenamefont {Vavrek}\ \emph {et~al.}(2017)\citenamefont {Vavrek},
  \citenamefont {Collins}, \citenamefont {Danagoulian}, \citenamefont
  {Henderson}, \citenamefont {Kemp}, \citenamefont {Lanza},\ and\ \citenamefont
  {Macdonald}}]{vavrek2017progress}%
  \BibitemOpen
  \bibfield  {author} {\bibinfo {author} {\bibfnamefont {J.~R.}\ \bibnamefont
  {Vavrek}}, \bibinfo {author} {\bibfnamefont {S.~J.}\ \bibnamefont {Collins}},
  \bibinfo {author} {\bibfnamefont {A.}~\bibnamefont {Danagoulian}}, \bibinfo
  {author} {\bibfnamefont {R.~S.}\ \bibnamefont {Henderson}}, \bibinfo {author}
  {\bibfnamefont {R.~S.}\ \bibnamefont {Kemp}}, \bibinfo {author}
  {\bibfnamefont {R.}~\bibnamefont {Lanza}}, \ and\ \bibinfo {author}
  {\bibfnamefont {R.}~\bibnamefont {Macdonald}},\ }in\ \href@noop {} {\emph
  {\bibinfo {booktitle} {Proc. 58th Annual INMM Meeting}}}\ (\bibinfo
  {organization} {Institute for Nuclear Materials Management},\ \bibinfo {year}
  {2017})\BibitemShut {NoStop}%
\bibitem [{\citenamefont {Kirkpatrick}\ and\ \citenamefont
  {Miyake}(1932)}]{gvm}%
  \BibitemOpen
  \bibfield  {author} {\bibinfo {author} {\bibfnamefont {P.}~\bibnamefont
  {Kirkpatrick}}\ and\ \bibinfo {author} {\bibfnamefont {I.}~\bibnamefont
  {Miyake}},\ }\href {\doibase 10.1063/1.1748828} {\bibfield  {journal}
  {\bibinfo  {journal} {Review of Scientific Instruments}\ }\textbf {\bibinfo
  {volume} {3}},\ \bibinfo {pages} {1} (\bibinfo {year} {1932})}\BibitemShut
  {NoStop}%
\bibitem [{\citenamefont {{Mirion Technologies (Canberra),
  Inc.}}(2017)}]{canberra2017labr}%
  \BibitemOpen
  \bibfield  {author} {\bibinfo {author} {\bibnamefont {{Mirion Technologies
  (Canberra), Inc.}}},\ }\href@noop {} {\enquote {\bibinfo {title}
  {{LABR-1.5x1.5 data sheet}},}\ } (\bibinfo {year} {2017}),\ \bibinfo {note}
  {retrieved from
  \url{http://www.canberra.com/products/detectors/pdf/LABR-SS-C39490.pdf} on
  Oct.~23, 2017.}\BibitemShut {Stop}%
\bibitem [{\citenamefont {Quarati}\ \emph {et~al.}(2013)\citenamefont {Quarati}
  \emph {et~al.}}]{QUARATI2013596}%
  \BibitemOpen
  \bibfield  {author} {\bibinfo {author} {\bibfnamefont {F.~G.~A.}\
  \bibnamefont {Quarati}} \emph {et~al.},\ }\href {\doibase
  https://doi.org/10.1016/j.nima.2013.08.005} {\bibfield  {journal} {\bibinfo
  {journal} {Nucl Instrum Methods Phys Res A}\ }\textbf {\bibinfo {volume}
  {729}},\ \bibinfo {pages} {596 } (\bibinfo {year} {2013})}\BibitemShut
  {NoStop}%
\bibitem [{\citenamefont {Agostinelli}\ \emph {et~al.}(2003)\citenamefont
  {Agostinelli} \emph {et~al.}}]{agostinelli_geant4}%
  \BibitemOpen
  \bibfield  {author} {\bibinfo {author} {\bibfnamefont {S.}~\bibnamefont
  {Agostinelli}} \emph {et~al.} (\bibinfo {collaboration} {GEANT4}),\ }\href
  {\doibase 10.1016/S0168-9002(03)01368-8} {\bibfield  {journal} {\bibinfo
  {journal} {Nucl. Instrum. Meth.}\ }\textbf {\bibinfo {volume} {A506}},\
  \bibinfo {pages} {250} (\bibinfo {year} {2003})}\BibitemShut {NoStop}%
\bibitem [{\citenamefont {Fetter}\ \emph
  {et~al.}(1990{\natexlab{b}})\citenamefont {Fetter}, \citenamefont {Cochran},
  \citenamefont {Grodzins}, \citenamefont {Lynch},\ and\ \citenamefont
  {Zucker}}]{ref:fetter1990gamma}%
  \BibitemOpen
  \bibfield  {author} {\bibinfo {author} {\bibfnamefont {S.}~\bibnamefont
  {Fetter}}, \bibinfo {author} {\bibfnamefont {T.~B.}\ \bibnamefont {Cochran}},
  \bibinfo {author} {\bibfnamefont {L.}~\bibnamefont {Grodzins}}, \bibinfo
  {author} {\bibfnamefont {H.~L.}\ \bibnamefont {Lynch}}, \ and\ \bibinfo
  {author} {\bibfnamefont {M.~S.}\ \bibnamefont {Zucker}},\ }\href@noop {}
  {\bibfield  {journal} {\bibinfo  {journal} {Science}\ }\textbf {\bibinfo
  {volume} {248}},\ \bibinfo {pages} {828} (\bibinfo {year}
  {1990}{\natexlab{b}})}\BibitemShut {NoStop}%
\end{thebibliography}%
\appendix
\renewcommand{\thesection}{S\arabic{section}}
\renewcommand{\theequation}{S\arabic{equation}}
\renewcommand{\thefigure}{S\arabic{figure}}
\renewcommand{\thetable}{S\arabic{table}}
\setcounter{figure}{0}
\setcounter{table}{0}
%
%
%
%
%
%

\section{Nuclear resonance fluorescence}\label{sec:si_nrf}
Nuclear resonance fluorescence (NRF) describes the X$(\gamma, \gamma')$X 
reaction in which a photon $\gamma$ is resonantly absorbed by the nucleus X and 
then re-emitted as the excited nucleus subsequently relaxes to its ground 
state~\cite{metzger1959resonance,kneissl1996structure}. Due to the 
discrete energy level structure of the nucleus, the probability that an incident 
photon of energy $E$ undergoes an NRF interaction is only significant if $E$ is 
approximately equal to one of the resonance energies $E_r$ of the nucleus, given 
by
\begin{align}
    E_r = E_\ell + \frac{E^2}{Mc^2}
\end{align}
where $E_\ell$ is the energy of a nuclear level and the latter term corrects for 
the recoil energy ($\sim$20~eV for U-238 and $E=2$~MeV) of the nucleus X with 
mass $M$. The probability of absorption by state $r$ is then given by the NRF 
cross section, which is most accurately described by a Doppler-broadened version of the
Lorentzian profile of Eq.~\ref{eq:sigmaNRFBW}:
\begin{align}\label{eq:sigmaNRF}
\begin{split}
    \sigma_{r}^\text{NRF}(E) &= 2\pi^{1/2} g_r \left( \frac{\hbar c}{E_r} 
\right)^2 \frac{b_{r,0}}{t^{1/2}} \int_{-\infty}^{+\infty} \exp\left[ 
-\frac{(x-y)^2}{4t}\right] \frac{dy}{1+y^2},
\end{split}
\end{align}
where
\begin{align}
    x \equiv 2(E-E_r)/\Gamma_r\\
    t \equiv (\Delta/\Gamma_r)^2,
\end{align}
$\Gamma_r$ is the intrinsic width of the excited state $r$, and
\begin{align}
    \Delta = E\sqrt{\frac{2k_B T}{Mc^2}}
\end{align}
is the Doppler-broadened width of the state at temperature $T$. For NRF lines of 
high-$Z$ isotopes, $\Gamma_r{\sim}1$--$100$~meV~\cite{nndc2015u238} while 
$\Delta{\sim}1$~eV. For greater accuracy, the $\Delta$ and thus $t$ may be evaluated 
using the effective temperature $T_\text{eff}$ instead of the physical temperature $T$ of the 
target~\cite{metzger1959resonance}. The $g_r$ is a statistical factor that 
accounts for the number of available spin states at the ground and resonant 
states:
\begin{align}
    g_r = \frac{2J_r+1}{2(2J_0+1)}
\end{align}
where $J_i$ for $i=\{0,r\}$ is the spin of the $i^\text{th}$ level. The branching ratio 
$b_{r,0}$ from the resonant state $r$ to ground also enters the calculation as 
$b_{r,0} \equiv \Gamma_{r,0}/\Gamma_r$, where $\Gamma_{r,0}$ is the partial 
width for the decay $r \to 0$ and $\sum_i \Gamma_{r,i} = \Gamma_r$. The 
$b_{r,i}$ therefore also give the probabilities of the resonant state $r$ 
decaying either directly to the ground state, emitting a photon of energy $E'= 
E_r$ (neglecting recoil), or through the intermediate state $i$, emitting a 
photon of energy $E' = E_r - E_i$.

Given Eq.~\ref{eq:sigmaNRF}, the NRF measurement described in the main paper can 
be described by a slightly simplified 1D model (see 
e.g.~Fig.~\ref{fig:schematic}) in which a parallel incident bremsstrahlung beam 
$\phi_0(E)$ is incident on a single-isotope rectangular slab warhead of 
thickness $D$. The transmitted flux $\phi_t(E)$ then interacts with a 
rectangular slab foil of thickness $X$ composed of the warhead isotope. In this 
case, the transmitted flux $\phi_t(E)$ through the warhead can be written as
\begin{align}\label{eq:phi_t}
    \phi_t(E) = \phi_0(E) \exp\left[ -D\left( \mu_\text{nr}(E) + 
\mu_\text{NRF}(E) \right) \right],
\end{align}
where the $\mu \equiv N \sigma$ terms denote linear attenuation coefficients if 
$D$ is expressed as a length, or mass attenuation coefficients if it is 
expressed as an areal density. This equation assumes that every NRF or 
non-resonant (`nr') interaction (e.g.~Compton scattering, pair production, etc.) 
results in the loss of forward-going flux at energy $E$. Because of the sharp 
$E$-dependence of $\mu_\text{NRF}(E)$, the forward-going flux is preferentially 
attenuated or `notched' at the resonance energies $E_r$ of the isotopes present 
in the warhead. The above assumption regarding photon losses can break down via 
a process known as `notch refill,' by which photons undergo small-angle Compton 
scattering to the resonance energies, thus replenishing the available flux in 
the notches and reducing the sensitivity of the measurement to the 
warhead~\cite{pruet2006detecting}. Since the notches are narrow, notch refill is 
only significant for relatively thick measurement targets (e.g.~a correction 
factor of $5\%$ for areal densities ${\sim}90$~g/cm$^2$~\cite{ref:quiter}) with 
many opportunities for downscatter.

For a single transition from $r\to 0$ (dropping subscripts for brevity), the 
double-differential NRF count rate induced by the transmitted flux $\phi_t(E)$ 
as observed by a single HPGe detector is
\begin{align}\label{eq:d2ndEdOmega}
\begin{split}
    \frac{d^2n}{dE d\Omega} &= \phi_t(E)\, b\, \mu_\text{NRF}(E) 
\frac{W(\theta)}{4\pi}  \frac{1-\exp\left[ -X \mu_\text{eff}(E,E',\theta) 
\right]}{\mu_\text{eff}(E,E',\theta)} \epsilon_\text{int}(E') P_f(E')
\end{split}
\end{align}
where $W(\theta)$ is the angular correlation function for successive gamma 
rays~\cite{hamilton1940directional} and $\epsilon_\text{int}(E')$ is the 
intrinsic peak efficiency of the HPGe detector. A high-$Z$ (typically Pb) filter 
may be placed between the foil and detector in order to preferentially attenuate 
low-energy photons and reduce detector dead time, in which case $P_f(E') < 1$ is 
the probability that an NRF photon of energy $E'$ will be transmitted through 
the filter. The effective attenuation coefficient
\begin{align}
    \mu_\text{eff}(E,E',\theta) \equiv \mu_\text{NRF}(E) + \mu_\text{nr}(E) + 
\frac{\mu_\text{nr}(E')}{\cos\theta}
\end{align}
accounts for attenuation in the foil of incoming photons (of energy $E$) via NRF 
and non-resonant processes as well as the attenuation of outgoing NRF photons 
(of energy $E'$) through the path at angle $\theta$ pointing to the detector. 
Integration of Eq.~\ref{eq:d2ndEdOmega} over all energies $E$ and the solid 
angle of the detector $\Omega$ then gives the predicted count rate for a single 
NRF peak as observed by the detector. The peak will appear not as a perfectly 
sharp emission line at $E'$, but as a Gaussian centered at $E'$ due to the 
imperfect resolution of the detector.

We note that the sharp $E$-dependence of $\mu_\text{NRF}(E)$ in the exponential 
terms of Eqs.~\ref{eq:phi_t} and \ref{eq:d2ndEdOmega} can substantially affect 
the predicted detected count rate~\cite[Fig.~3.25]{quiter2010thesis}: while the 
Doppler-broadened Lorentzian profile of Eq.~\ref{eq:sigmaNRF} is the most 
accurate, a Gaussian approximation~\cite{metzger1959resonance} to the cross 
section is often sufficient. The rectangular cross section approximation---a 
constant value of cross section over an energy range on the order of $1$~eV such 
that the integral of Eq.~\ref{eq:sigmaNRF} over $E$ (the `integrated cross 
section') is preserved---is only accurate to about $20\%$ and should be avoided 
unless computational efficiency is required at the expense of accuracy.

\section{Experimental Methods}\label{sec:si_exp}

\subsection{Data acquisition}
%
%

Data acquisition (DAQ) was accomplished using a Canberra Lynx Digital Signal 
Analyzer (DSA) connected to each HPGe detector~\cite{ortec_gem}. Instead of using the standard 
Genie2K acquisition software, the three detectors and DAQs were controlled 
simultaneously using the custom-written Python Readout with Lynx for Physical 
Cryptography (PROLyPhyC) wrapper classes sitting atop the Lynx Software 
Development Kit (SDK). Events were recorded in pulse-height analysis (PHA) mode, 
resulting in a 32768-channel pulse height spectrum produced for each detector at 
the end of each acquisition time. To guard against data corruption due to beam 
instability, acquisition times were set to five minutes (real time); each 
measurement therefore consisted of $\sim$10 such acquisition periods summed 
together using the procedure described in Section \ref{sec:si_data}. The raw 
pulse height spectra were converted to energy (deposition) spectra using linear 
calibrations of channel number vs energy using Cs-137 ($0.662$~MeV) and Co-60 
($1.172$, $1.333$~MeV) check sources taken before and throughout the week of 
experiments.

The integrated beam charge over the course of an acquisition period was 
determined by using a Keithley Model 614 Electrometer to measure the 
beam-induced current from the radiator to ground. The analog output of the 
electrometer was digitized by a Measurement Computing Model USB-201 
analogue-to-digital converter at a sample rate of 1~kHz, and read to a plain 
text file on a laptop for persistent storage. The average current over the 
acquisition time was then computed for use in Eq.~\ref{eq:live_charge_norm} and 
compared against the display of the electrometer throughout the run for 
consistency.

\begin{figure}[thb]
    \centering
    \includegraphics[width=0.6\columnwidth]{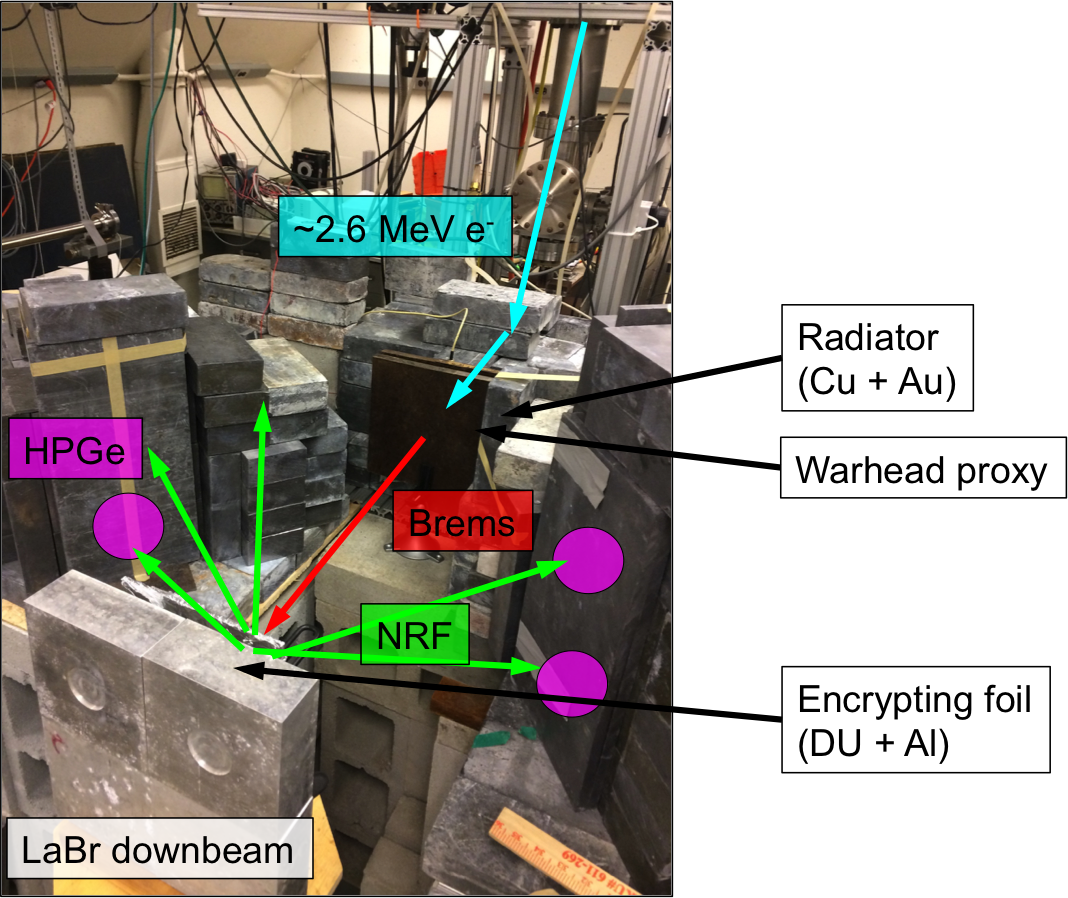}
    \caption{Annotated photograph of the target geometry.}
    \label{fig:setup_photo}
\end{figure}

\begin{figure}[thb]
    \centering
    \includegraphics[width=0.6\columnwidth,angle=270]{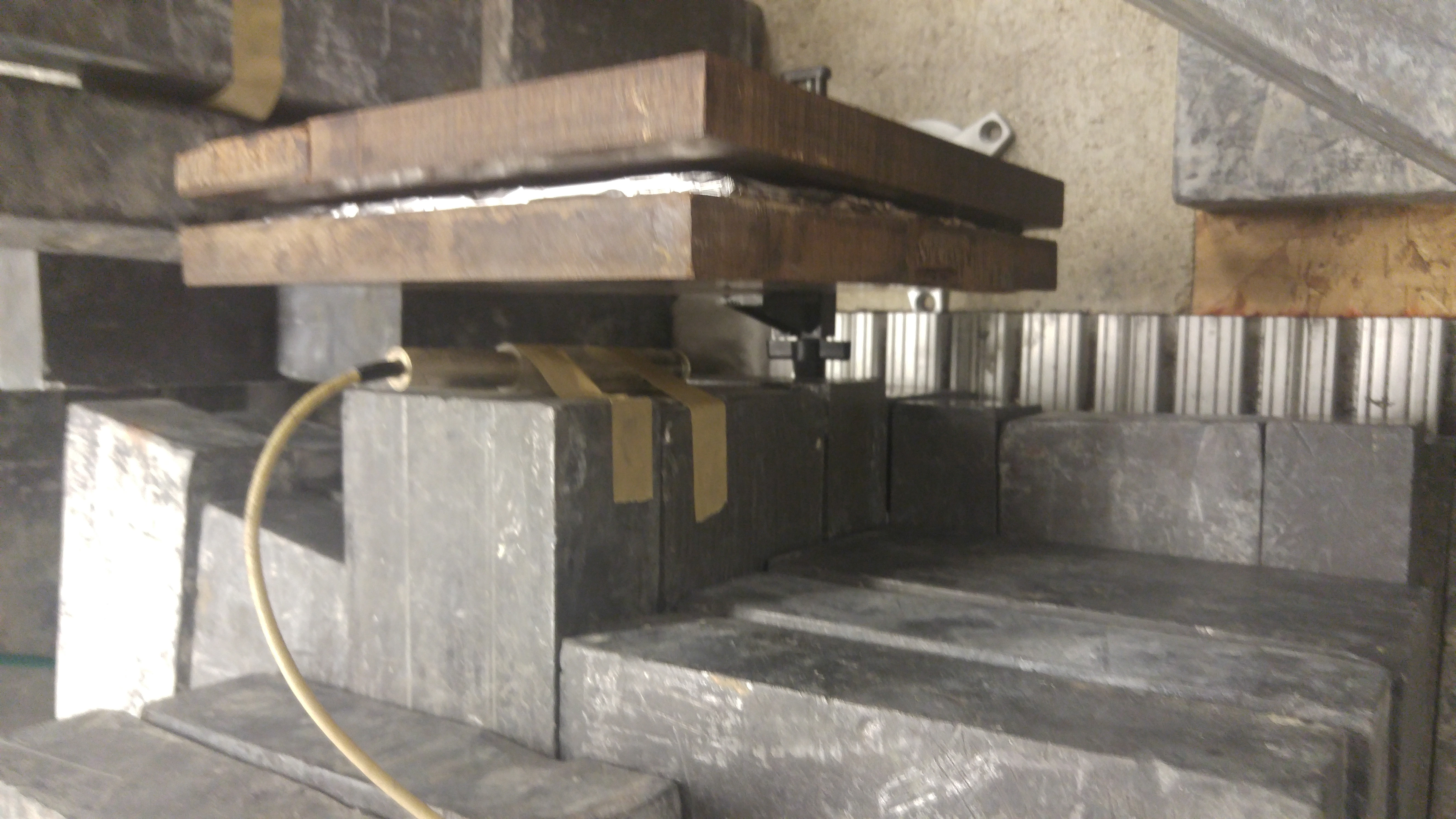}
    \caption{Close-up of template I near the collimator exit. The cylinder 
affixed to the end of the collimator is a gas ionization chamber used for beam 
tuning and monitoring.}
    \label{fig:genuine_photo}
\end{figure}

%
%

%
%

\subsection{Beam characterization and stability}\label{sec:si_beam_char}

The stability and reproducibility of the electron beam settings (most notably, 
the beam energy, current, and position relative to the gold radiator foil) 
directly affects the validity of comparisons between template/candidate 
scenarios, especially when the integrated beam charge is used to normalize 
measurements.  In particular, preliminary experiments that tested elements of 
the physical cryptographic protocol prior to the work reported here indicated 
that the absolute rate of NRF photon detection was lower than expected from 
simulation and analytic calculations by a factor of $1.5$--$2$ 
\cite{vavrek2017progress}.  A number of possible explanations for this 
observation were explored, and uncertainties in the electron beam position, 
emittance, and energy were identified as the most likely causes of the 
discrepancy.  While knowledge of the \textit{absolute} bremsstrahlung flux is 
not required to perform the relative spectral comparisons between template and 
candidate warheads, any temporal variance in the flux could make such 
comparisons invalid.  Due to this, several operational procedures and 
diagnostics were implemented to complement and enhance the existing HVRL beam 
diagnostics.  The results of these diagnostics are presented in this section,
demonstrating that while variations in the beam may have affected previous
experiments, the beam conditions were well-understood and constrained for
the data presented in this work due to the improvements.

%
%

\subsubsection{Electron beam energy}

\begin{figure}[thb]
    \centering
    \includegraphics[width=0.8\columnwidth]{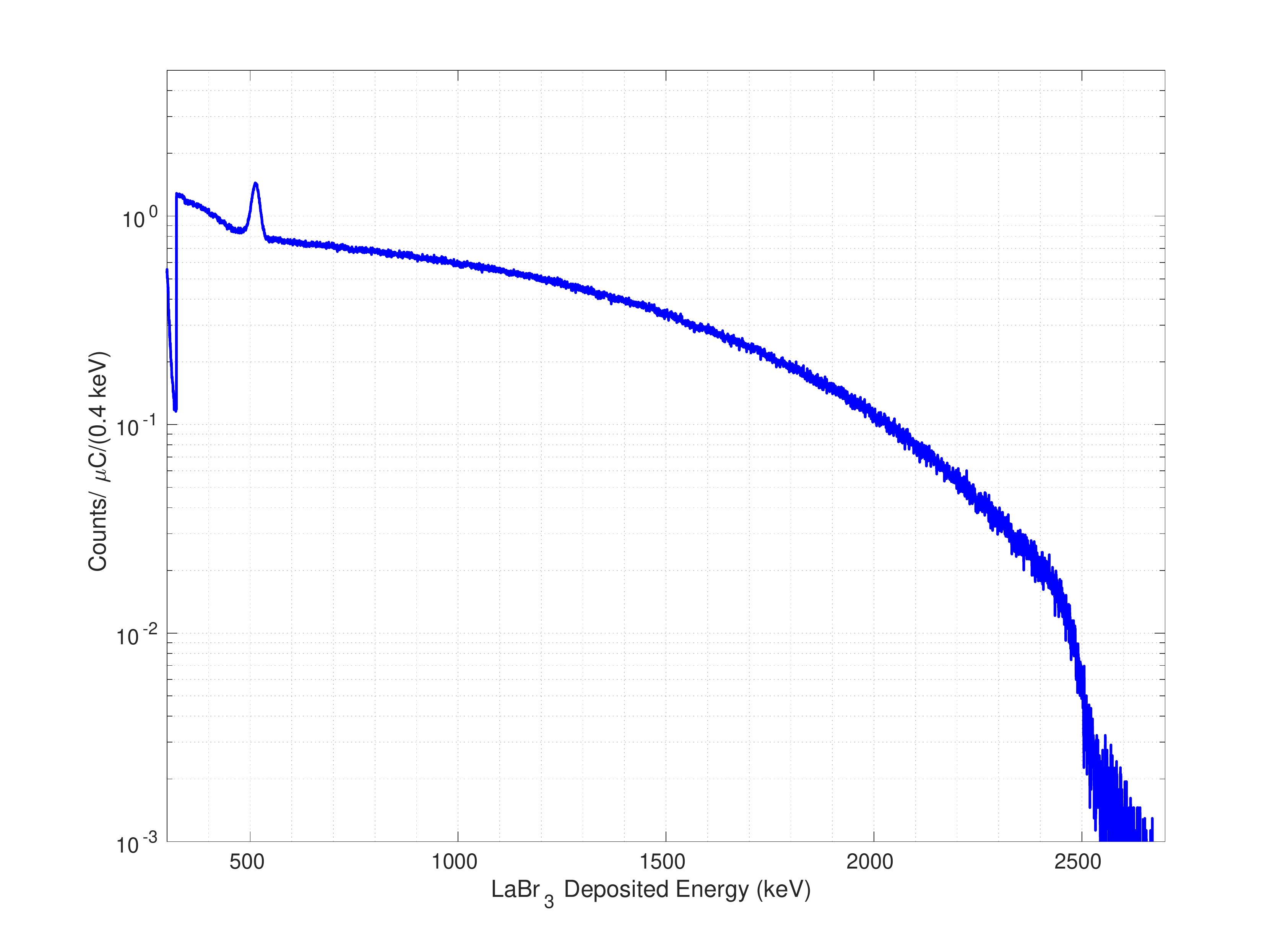}
    \caption{Calibrated spectrum of the bremsstrahlung beam recorded by the 
LaBr$_3$ detector after transmission through the encryption foil and the 6-inch 
Pb filter shielding the scintillator at a nominal beam energy of 2.6 MeV. In this and all subsequent Figures, error bars are $\pm$1 SD.}
    \label{fig:typspect}
\end{figure}

\begin{figure}[thb]
    \centering
    \includegraphics[width=0.8\columnwidth]{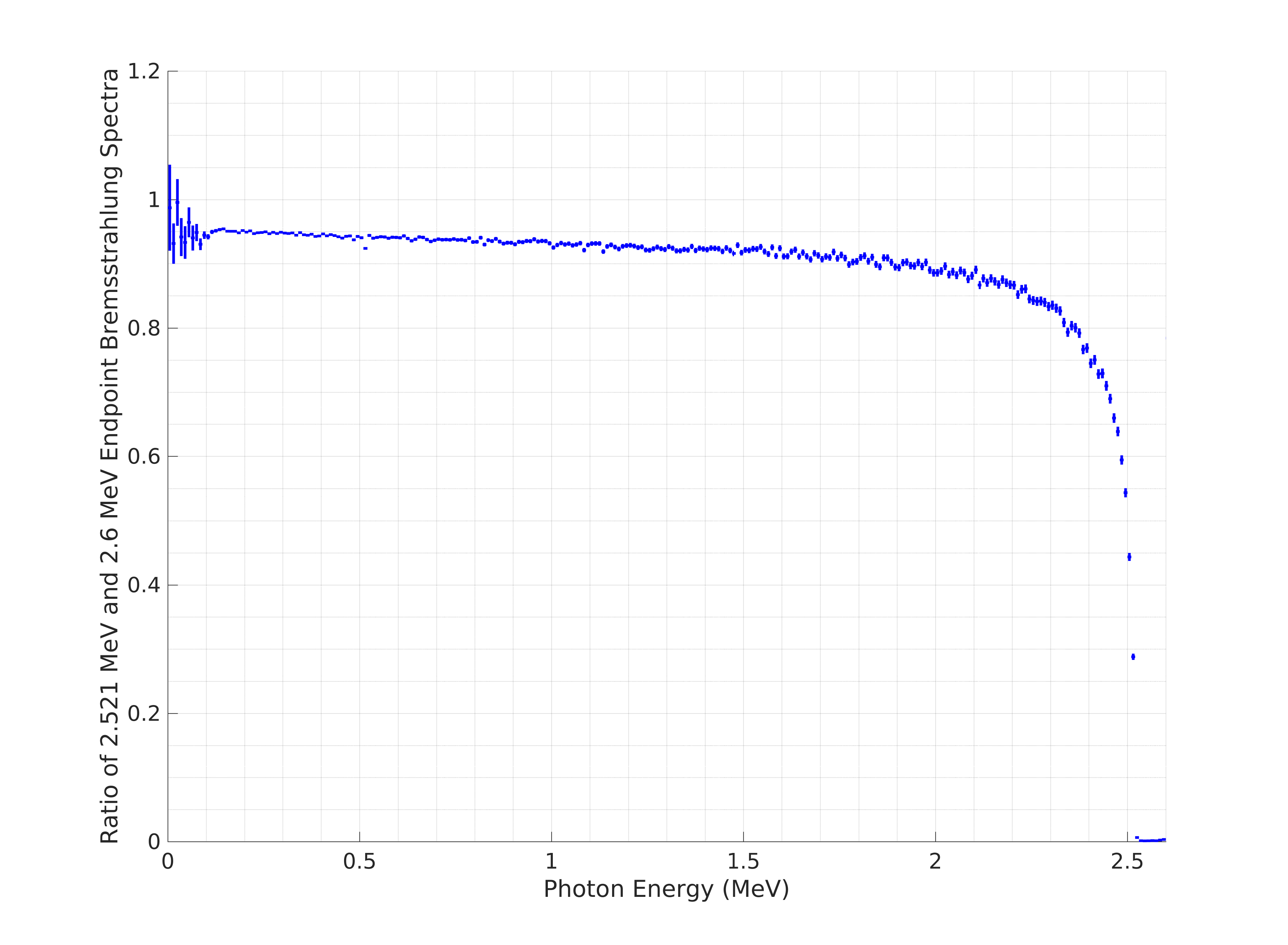}
    \caption{Ratio of the simulated 2.521 MeV endpoint bremsstrahlung spectra to the simulated 2.6 MeV endpoint spectra, showing the order
             10\% deficit of photons in the former case relative to the latter in the region of interest around 2 MeV.}
    \label{fig:bremrat}
\end{figure}

The electron beam kinetic energy was chosen as 2.6 MeV as a compromise of 
several competing factors.  An ideal beam for this application maximizes the 
number of photons at the specific energies of the NRF lines of interest while 
minimizing photons at other energies.  Photons above the NRF energies may 
undergo various physical processes that may cause them to scatter into the 
detectors resulting in background counts in the region of the spectrum near the 
NRF energies and additionally contribute to the notch refill effect discussed in 
Section \ref{sec:si_nrf}. Below the NRF energies, photons contribute to pile-up 
effects in detectors and add to the radiation dose to which inspected objects 
are exposed.  To balance these effects when studying NRF lines, it is most 
effective to choose an endpoint energy a few hundred keV above the NRF energies. 
For a photon source produced by the bremsstrahlung of electrons, the number of 
photons rapidly decreases as a function of energy with no photons produced above 
the energy of the incident electrons, as visible in the spectra shown in Fig. 
\ref{fig:typspect}.  Due to this sharp drop-off in the spectrum, however, the 
total flux of photons at the NRF energies depends strongly on the precise 
location of the endpoint.  This is illustrated in Fig.~\ref{fig:bremrat}, which 
shows the ratio of the forward bremsstrahlung fluxes of electron beams of 
nominal (2.6 MeV) energy and of energy below nominal (2.521 MeV).  This 
$\sim$3\% change in the beam energy results in $\gtrsim$10\% change at the NRF 
lines energies, which is further magnified by the even greater reduction at 
higher energies (since these photons can downscatter within the mock warhead 
and/or foil to add to the flux).  While the absolute flux of the bremsstrahlung 
beam is not required for the comparative measurements presented here, this 
effect necessitates establishing that the beam energy was consistent between 
measurements.  The HVRL electron beam energy was set using a generating 
voltmeter (GVM), which measured the potential across the accelerator terminals 
\cite{gvm}.  When used for this purpose, however, GVMs require regular, 
independent calibration to the actual electron beam energy, a process that had 
not been conducted for the HVRL beam for some time prior to the experiments 
described in this work.  Additionally, since the GVM reading was not recorded 
throughout the run (so as to monitor its fluctuations), it is critical to 
establish the stability of the beam energy between the different measurements.

\begin{figure}[thb]
    \centering
    \includegraphics[width=1.0\columnwidth]{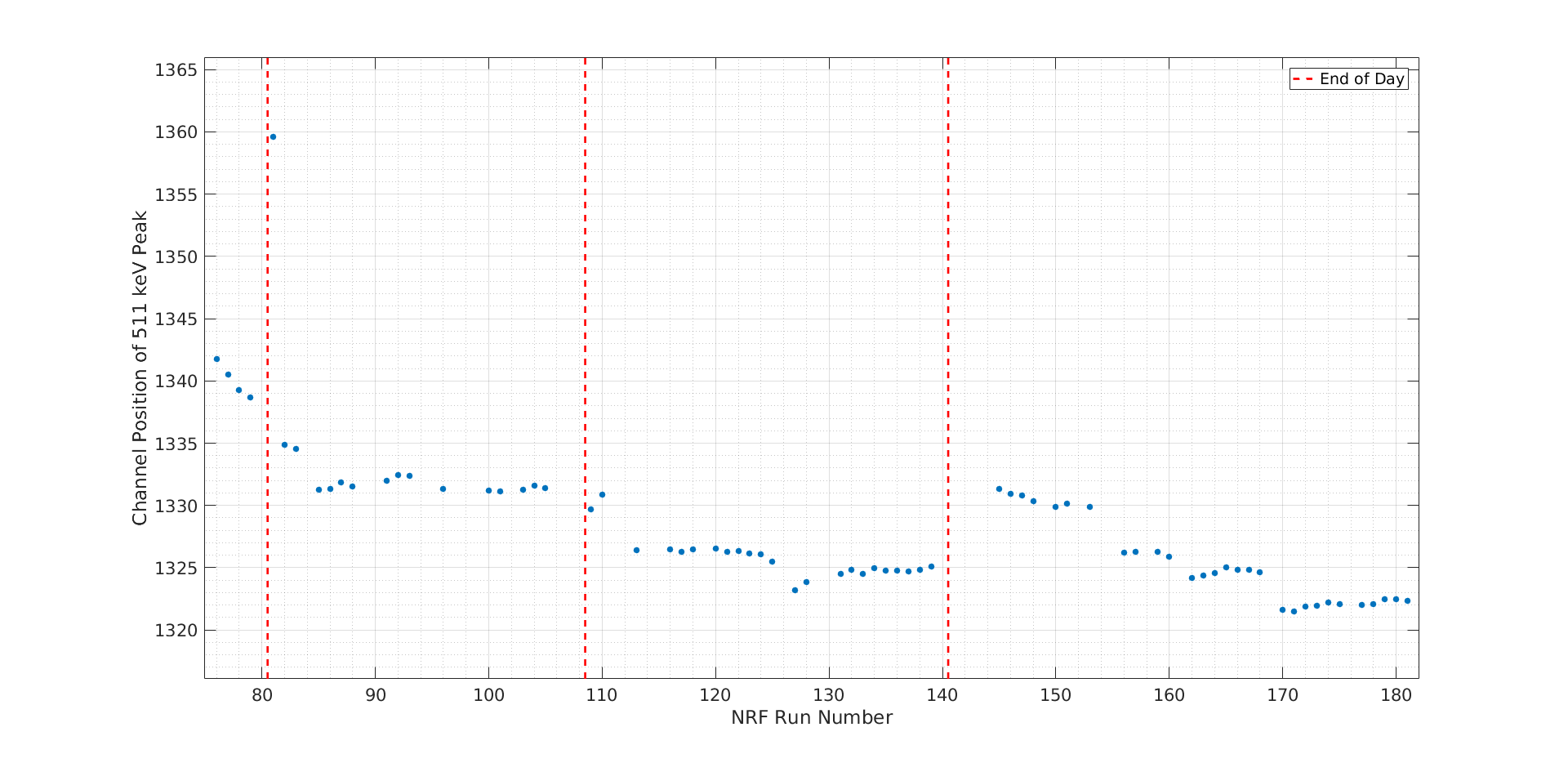}
    \caption{Position of the 511 keV pair production peak in the LaBr$_3$ over 
the course of the entire experimental run for the data presented in this paper.  
With the exception of Run \#81, the gain of the LaBr$_3$ detector was stable to 
within 1.5\%.  Dashed lines indicate gaps between days of operation.}
    \label{fig:labrgain}
\end{figure}

To measure the electron beam energy, the LaBr$_3$ scintillator spectra of the 
bremsstrahlung photons for each data run were examined.  The LaBr$_3$ detector~\cite{canberra2017labr}
is especially well-suited for examining spectral features in the vicinity of 
2.0--2.7 MeV due to the presence of intrinsic spectral lines in this region that 
are due to alpha decays from the decay chain of Ac-227, which is a contaminant 
in LaBr$_3$ due to the chemical similarities of La and Ac \cite{QUARATI2013596}. 
 The electronic-equivalent energy depositions from several of these decays allow 
a precise calibration of the ADC-photon energy calibration of the detector in 
this region.  Following the experimental run, a long sample of the intrinsic 
spectrum of the detector was collected to provide a precise energy calibration.  
To account for a possible shift in the gain of the detector over the course of 
the experimental run, the ADC channel position corresponding to the 511 keV peak 
(caused by the plentiful $e^+/e^-$ pair production interactions caused by 
photons with energy greater than 1022 keV in the bremsstrahlung beam) was 
determined for each of the data runs as a measure of the shift in gain and is 
shown in Fig.~\ref{fig:labrgain}.  The shift in this peak position relative to 
the data taken immediately before the intrinsic calibration run was used to 
correct the gain drift for each spectrum to calibrate the individual spectra. 
Additionally, pulse shape discrimination was utilized to exclude pile-up events 
(in which two photons contributed to a single count in the spectrum).  Since 
such events contribute relatively significantly to the high-energy end of the 
spectrum, rejecting them is necessary to sharply reconstruct features such as 
the spectral endpoint.

\begin{figure}[thb]
    \centering
    \includegraphics[width=1.0\columnwidth]{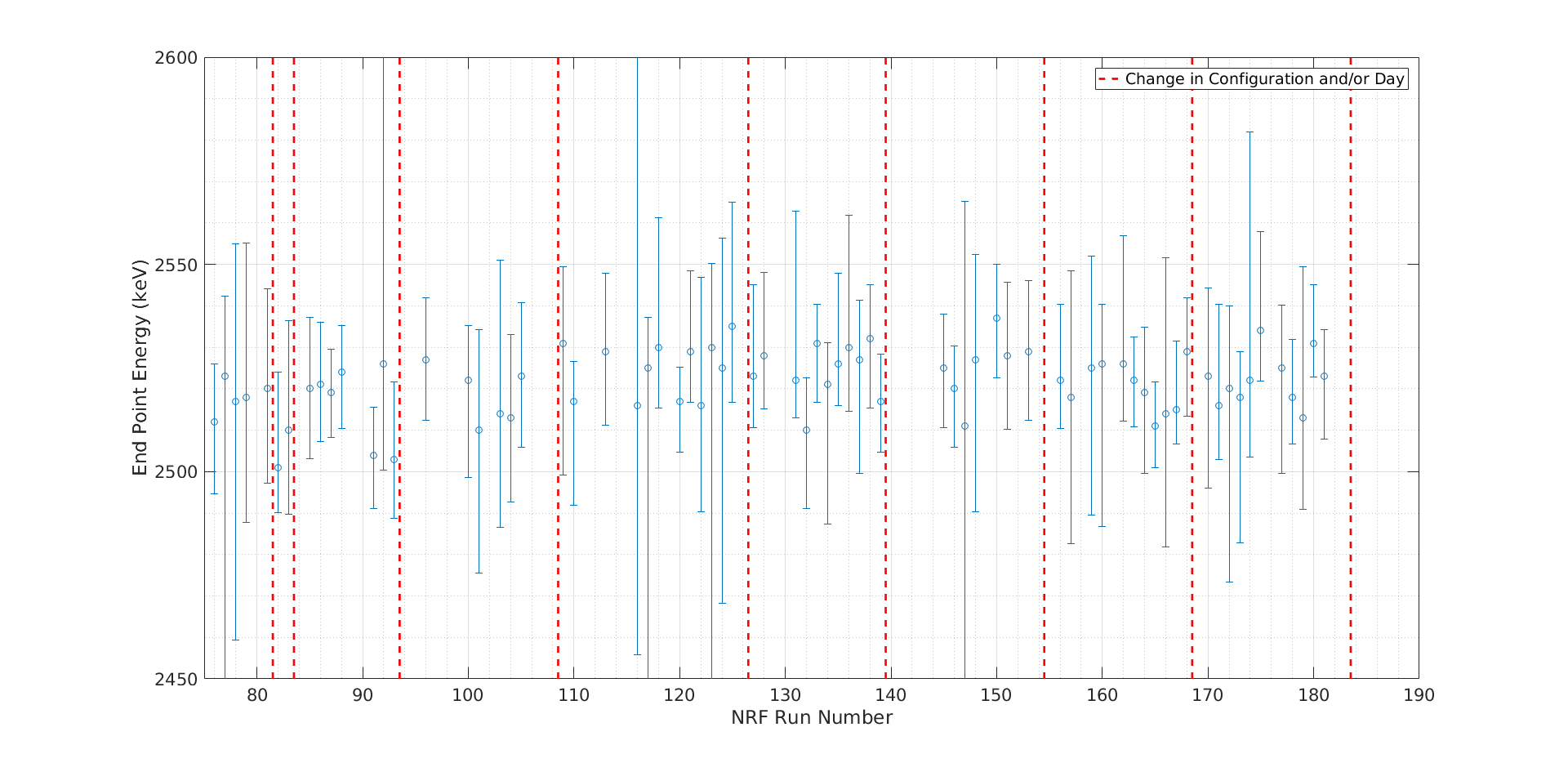}
    \caption{Extracted electron beam energy for each of the production data 
runs.  While below the nominal value of 2.6 MeV, the endpoint was consistent to 
within uncertainties for the entire data-taking period.}
    \label{fig:endpoint}
\end{figure}

\begin{figure}[thb]
    \centering
    \includegraphics[width=1.0\columnwidth]{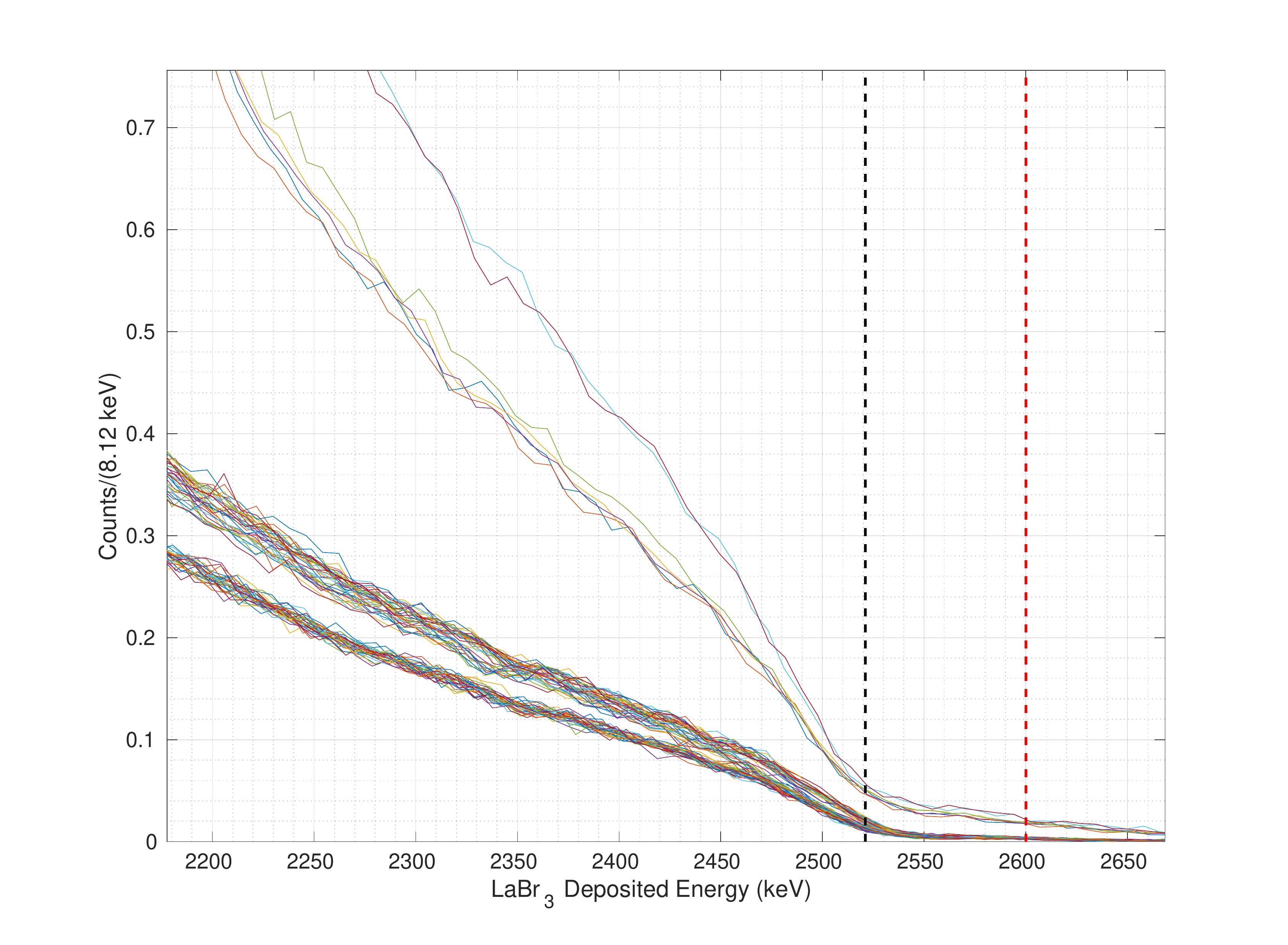}
    \caption{Overlain calibrated and pile-up rejected spectra from the LaBr$_3$ detector for all data runs, showing the difference between
    the extracted endpoint at 2521 keV (black dashed line) and the nominal endpoint at 2600 keV (red dashed line).}
    \label{fig:enddiff}
\end{figure}

For each calibrated spectrum, the point at which the second derivative of the 
spectrum was maximal (determined numerically) was found, indicating the position 
at which the rapidly decreasing bremsstrahlung spectrum met the relatively 
flatter background above the endpoint, thus indicating the energy of the 
incident electron beam.  Fig. \ref{fig:enddiff} shows the shape of the spectrum
endpoint for each data run, illustrating the fact that the rapid drop-off in the bremsstrahlung
spectrum occurred at a lower energy than the nominal 2.6 MeV endpoint.
Fig.~\ref{fig:endpoint} shows the results of the fit
determination for each of the data runs.  This procedure contributes 4 keV 
systematic uncertainty to the overall determination of the endpoint, while the 
gain drift conservatively contributes another $\sim$0.5\% uncertainty (reduced 
from the total drift by the correction described).  Averaging the results of the 
individual electron beam energy extractions results in a beam energy 
determination of $(2.521\pm0.015)$ MeV, lower than the nominally determined beam 
energy.  As shown in Fig.~\ref{fig:endpoint}, however, the endpoint energy was 
very stable to within the quoted uncertainty over the entire experimental run, 
demonstrating that variations in the beam energy did not systematically affect 
any comparisons between template and candidate proxy warheads.  Understanding 
this systemic offset, however, is critical for any future analyses that require 
knowledge of the absolute photon flux.

\subsubsection{Electron beam position}

The spectrum of bremsstrahlung photons emitted from a radiator depends 
significantly on the geometry and materials of the radiator, and thus may also 
depend on the position at which the electrons are incident on the radiator.  In 
particular, the number of photons generated near the endpoint energy is 
maximized by ensuring that the photons first strike the highest atomic number 
material (in this case the gold of the 126 {\textmu}m foil) prior to losing 
energy through interactions with other materials.  If the position of the 
electron beam deviates from the center of the foil or if the electron beam has a 
significant width beyond the 0.5 cm radius of the gold foil, the bremsstrahlung 
photon spectrum is altered and any inconsistencies in these parameters over the 
course of the experimental run could induce differences between the different 
proxy warhead tests.

\begin{figure}[thb]
    \centering
    \includegraphics[width=0.6\columnwidth]{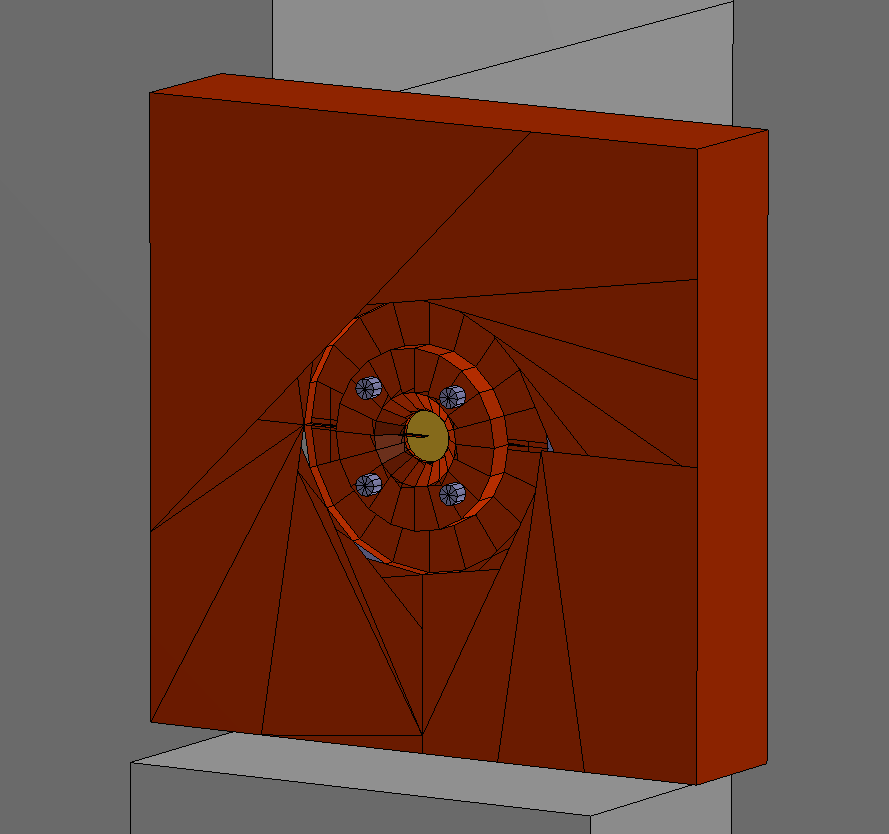}
    \caption{Visualization of the solid model of the bremsstrahlung radiator 
used for simulated beam studies.  The exposed gold foil in the center had a 
diameter of 1 cm.}
    \label{fig:radgeo}
\end{figure}

\begin{figure}[thb]
    \centering
    \includegraphics[width=1.0\columnwidth]{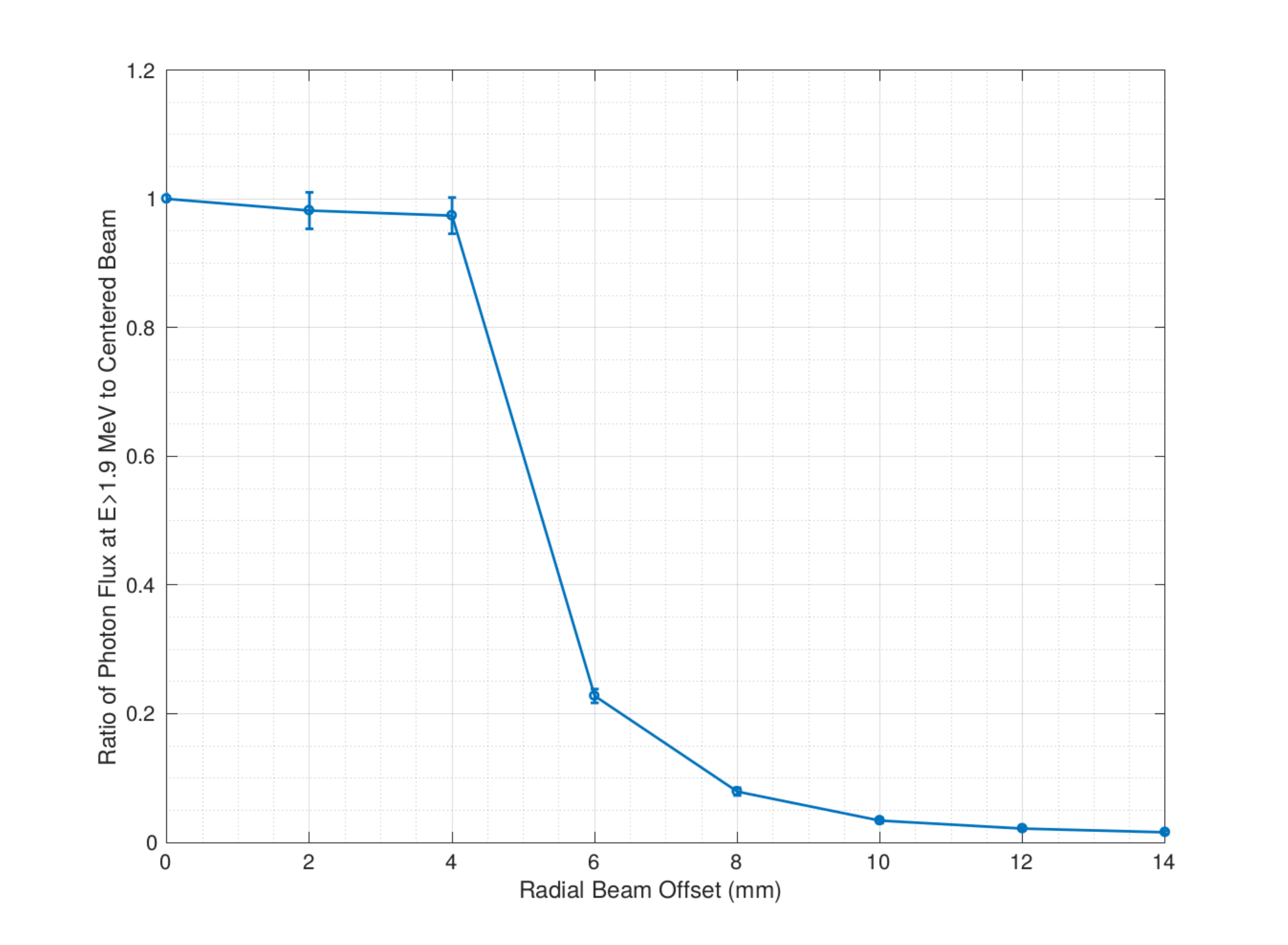}
    \caption{Simulated effect of drift in the beam position on the flux of 
photons with $E>1.9$ MeV incident on the mock warhead.}
    \label{fig:beam_offset}
\end{figure}

To study possible magnitude of this effect, the effect of beam wander on the 
bremsstrahlung spectrum was simulated using the Geant4 toolkit 
\cite{agostinelli_geant4}.  In this simulation, the geometry of the 
bremsstrahlung radiator and collimator were modeled in detail based on 
experimental survey of the objects, shown in Fig.~\ref{fig:radgeo}, and electron 
beams of energy 2.6 MeV were simulated incident upon the radiator at different 
positions.  The simulated beams were infinitely narrow and incident normal to 
the face of the radiator.  The incident position of the beam was varied radially 
from the center of the gold foil ($r=0$ mm) to beyond the foil radius so that 
the electrons were directly incident on the copper frame ($r=14$ mm).  For each 
beam position, the number of bremsstrahlung photons incident on the mock warhead 
target (i.e., beyond the collimator) was counted and compared to the number 
generated when the beam was centered at the same electron current.  The results 
of this study are shown in Fig.~\ref{fig:beam_offset}, plotted as the ratio of 
the number photons generated above 1.9 MeV (i.e., in the NRF region of interest) 
for a given beam position to the number generated with the beam on center.  The 
simulation shows that while the beam remains on the gold foil ($r<5$ mm) the 
number of high energy photons remains within a few percent of the ideal value.  
For $r>5$ mm, the high energy photon count drops precipitously.  Thus, as long 
as the beam remains on the foil throughout the experiments, the beam position 
uncertainty contributes a systematic uncertainty of at most $\lesssim$1\% to the 
comparisons of different mock warheads.  Large deviations in beam position, 
however, would have a significant effect on the data by greatly reducing the 
number of photons available for the NRF interactions.

During the experiments described in this work, there existed no means of 
concurrently monitoring the beam position and width.  Prior to these experiments 
however, the HVRL electron beam was imaged using a beryllium oxide screen as 
part of the experiments conducted by another user of the facility.  Using this 
imaging system, the electron beam focusing elements were tuned to minimize the 
beam diameter and maximize its positional stability for the 2.6 MeV beam energy 
setting required for the experiments used in this paper.  The BeO screen was 
placed 200 cm from the dipole magnet (located at the ``bend'' of the $e^-$ beam 
shown Fig. \ref{fig:setup_photo}).  At this distance, the electron beam diameter 
could be held to smaller than the 5 cm$\times$5 cm screen and stable in position 
to within a few millimeters (see the video included in the supplemental materials,
courtesy of C.S. Epstein).  Given that the radiator foil was 
located approximately 20 cm from the dipole bend, and that the beam is maximally 
focused when exiting the dipole, it is likely that the beam was well confined to 
the gold foil throughout the experimental run and thus the contribution to the 
uncertainty on the NRF lines measurements from beam wander is limited to 
$\sim$1\%.  Since this uncertainty is negligible compared to the statistical 
uncertainty of the data, it is neglected in the analysis.

%
%

\subsubsection{Beam conditioning}

Due to the age and the electrostatic mechanism of the HVRL electron accelerator, 
the accelerator had to be regularly `conditioned' by running an incoherent 
electron plasma (as opposed to a coherent beam) through the beamline to burn off 
contaminants.  Failure to regularly condition the beam would lead to deviations 
in the beam current and energy, invalidating any data collected until the beam 
was reconditioned.  Care was taken to regularly condition the beam and monitor 
its stability, and any data taken during periods in which a significant beam 
parameter deviated from the nominal value was excluded from the main analyses.  
Deviations were observed as uncontrolled shifts in the measured electron beam 
current, shifts in the measured terminal voltage, or unexplained changes or 
time-variance in the spectra from the HPGe and LaBr$_3$ detectors (which were 
monitored online).  During the experimental run in which the data presented here 
was collected, typically $30$ minutes of conditioning were required for every 
four hours of run time.  This frequency would have been higher were it not for 
the beam effectively being conditioned by another group's experiment during the 
previous week (during which the beam imaging and tuning described in the 
previous section was conducted).  It should be noted that a 
modern, dedicated accelerator facility for the purpose of an implemented weapons 
verification program would not face such limitations.

\subsubsection{Summary of beam effects}

The combination of this flux reduction due to the beam energy, past lack of 
constraint
on wandering of the beam condition, and less stable beam conditions in prior 
experiments is hypothesized to account for the factor of $1.5$--$2$ ratio of 
predicted over observed absolute NRF rates in preliminary 
experiments~\cite{vavrek2017progress}. The beam diagnostics used for the 
experimental campaign presented here indicate that these issues were, in 
general, rectified for the data presented.  In particular, these analyses 
indicate that the beam conditions were very stable.  Thus, while there may be 
remaining uncertainties on the absolute bremsstrahlung flux, this flux was 
consistent and thus relative comparisons between different mock warheads are not 
subject to significant uncertainties
from the beam conditions. Such cancellation of consistent, systematic 
uncertainties is an inherent advantage of any template verification system.

\section{Analysis of data from multiple detectors}\label{sec:si_data}
\subsection{Data unification}
For each measured object, data from multiple acquisition periods and three 
separate detectors is combined into a single spectrum in order to improve the 
signal-to-noise ratio. First, the count rate $r^{d}_i$ (counts per live second 
per {\textmu}A) in the $i^\text{th}$ bin of detector $d$'s spectrum is the 
live-charge-normalized sum of bin contents (i.e.~raw counts) $c^{d}_{ij}$ in 
each of the $j$ runs:
\begin{align}\label{eq:live_charge_norm}
    r^{d}_{i} = \frac{\sum_j c^{d}_{ij}}{\sum_j t^{d}_{\ell,j} I_{b,j}}
\end{align}
where $I_{b,j}$ is the average beam current recorded in run $j$, and the live 
time $t^{d}_{\ell,j}$ is computed from the real time $t^{d}_{r,j} = 300$~s and 
the detector dead time fraction $f^{d}_{\text{dt},j}$ (as reported by each 
detector's Lynx DSA) as
\begin{align}
    t^{d}_{\ell,j} = t^{d}_{r,j} (1-f^{d}_{\text{dt},j}).
\end{align}
To build a meaningful sum across the three detectors $d$, each histogram of 
rates $r^d$ must have an equal number of bins and locations of bin centers. This 
is difficult to achieve in practice, however, since each detector has a 
different calibration (depending on its gain and unique response function) for 
converting from a bin number in the range $1$-$32768$ to energy deposition in 
MeV. As a result, the calibrated bin widths and bin centers, in general, differ 
among the detectors. This is solved in post-processing by a combination of 
recalibration and histogram interpolation. As a common starting point, each 
histogram $r^d$ is linearly recalibrated using its peaks at $0.511$~MeV (pair 
production), $1.001$~MeV (U-238 passive signature), and $2.212$~MeV (Al-27 NRF 
emission), all of which are prominent in the beam-on spectrum. A new histogram 
$\bar{r}^{d}$ with $3000$ $1$~keV-wide bins between $0$ and $3$~MeV is then 
generated by interpolation of $r^d$, and scaled by the ratio of new to old bin 
widths in order to keep constant the differential counts per unit energy. The 
bin errors $\delta\bar{r}^{d}_i$ are finally recomputed under Poisson statistics 
as
\begin{align}
    \delta \bar{r}^{d}_i = \sqrt{\frac{\bar{r}^{d}_i}{\sum_j t^{d}_{\ell,j} 
I_{b,j}}}
\end{align}
which just amounts to reverting the live-charge-scaled spectra to count spectra, 
computing the bin error as the square root of the (interpolated) counts, and 
then re-dividing by the live charge. The detector-summed spectrum is then just 
the bin-by-bin sum $\bar{r}_i = \sum_d \bar{r}^d_i$. Note: the peak resolutions 
of $\sim$0.05\% ($\sigma$) at $\sim$2.2~MeV in the spectrum $\bar{r}$ (after 
processing) were verified to be consistent with those in the individual spectra 
$r^d$ (before processing).

\subsection{Fitting procedure}
\begin{figure}[thb]
    \centering
    \includegraphics[width=0.9\columnwidth]{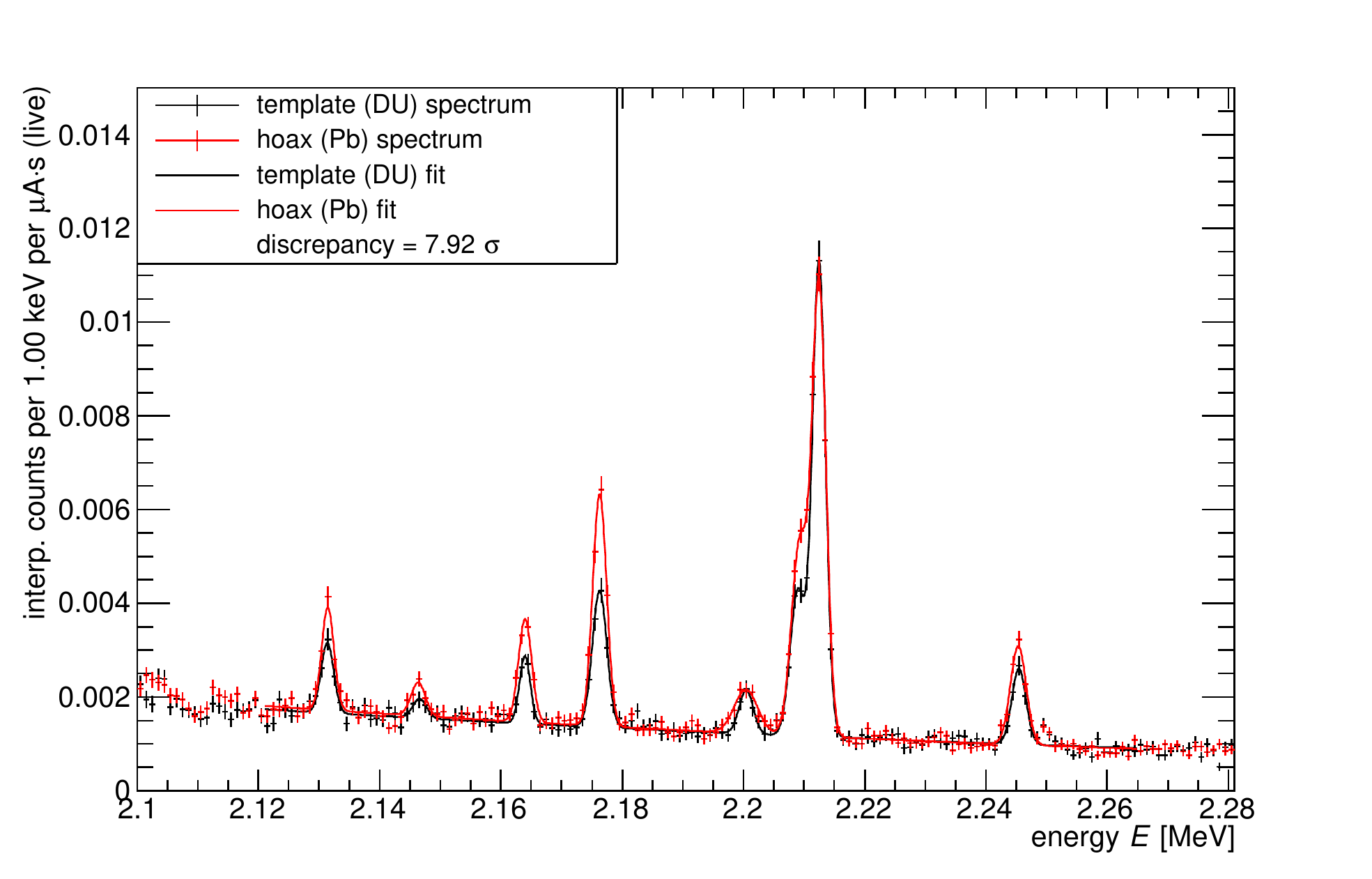}
    \caption{Summed spectra (points) and fits (curves) in the template I (black) 
vs hoax Ia (red) verification measurement.}
    \label{fig:I_vs_Ia}
\end{figure}

\begin{figure}[thb]
    \centering
    \includegraphics[width=0.9\columnwidth]{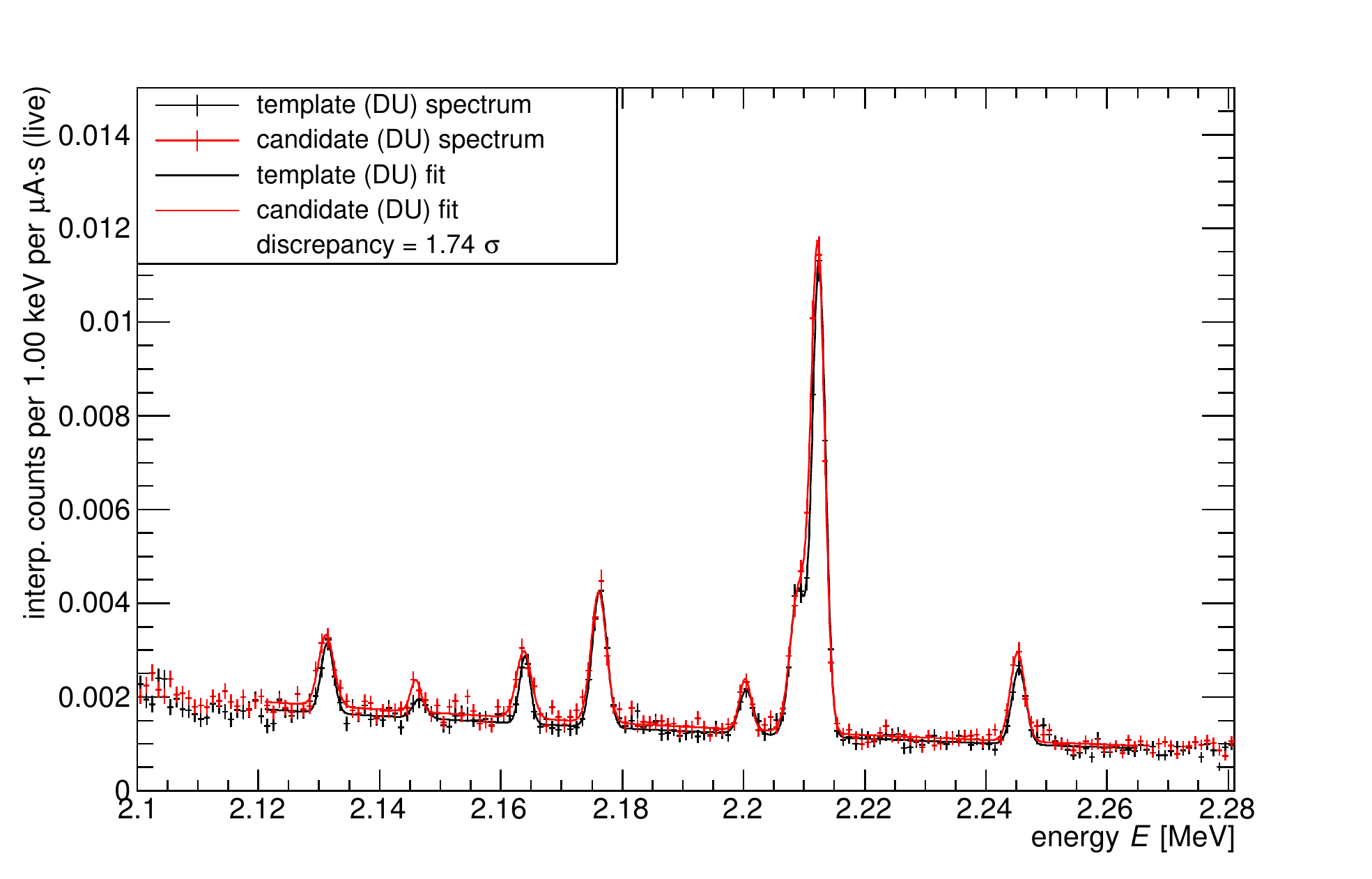}
    \caption{Summed spectra (points) and fits (curves) in the template I (black) 
vs genuine candidate Ig (red) verification measurement.}
    \label{fig:I_vs_Ig}
\end{figure}

\begin{figure}[thb]
    \centering
    \includegraphics[width=0.9\columnwidth]{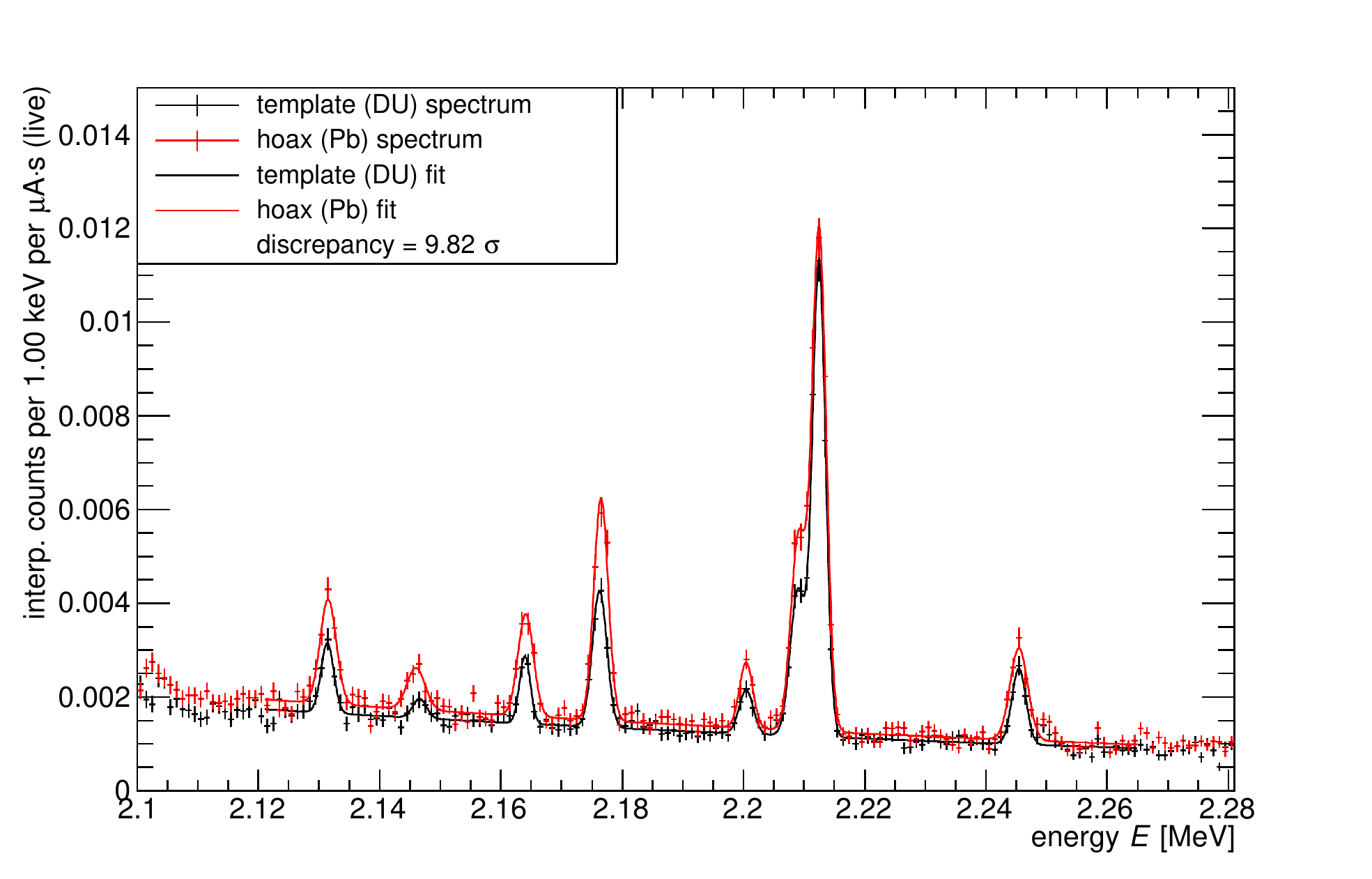}
    \caption{Summed spectra (points) and fits (curves) in the template I (black) 
vs hoax Ib (red) verification measurement.}
    \label{fig:I_vs_Ib}
\end{figure}

\begin{figure}[thb]
    \centering
    \includegraphics[width=0.9\columnwidth]{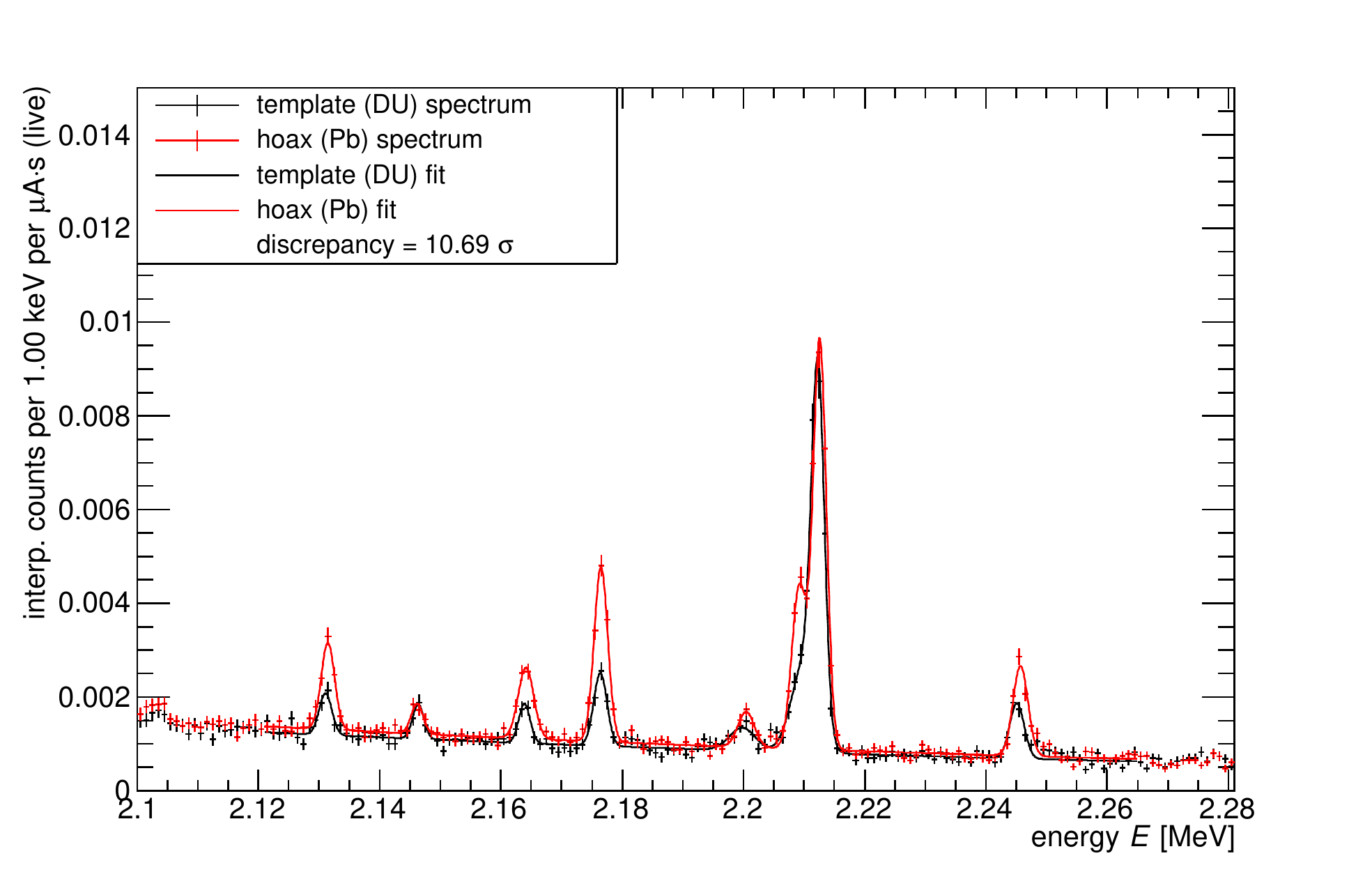}
    \caption{Summed spectra (points) and fits (curves) in the template II 
(black) vs hoax IIc (red) verification measurement.}
    \label{fig:II_vs_IIc}
\end{figure}

\begin{figure}[thb]
    \centering
    \includegraphics[width=0.9\columnwidth]{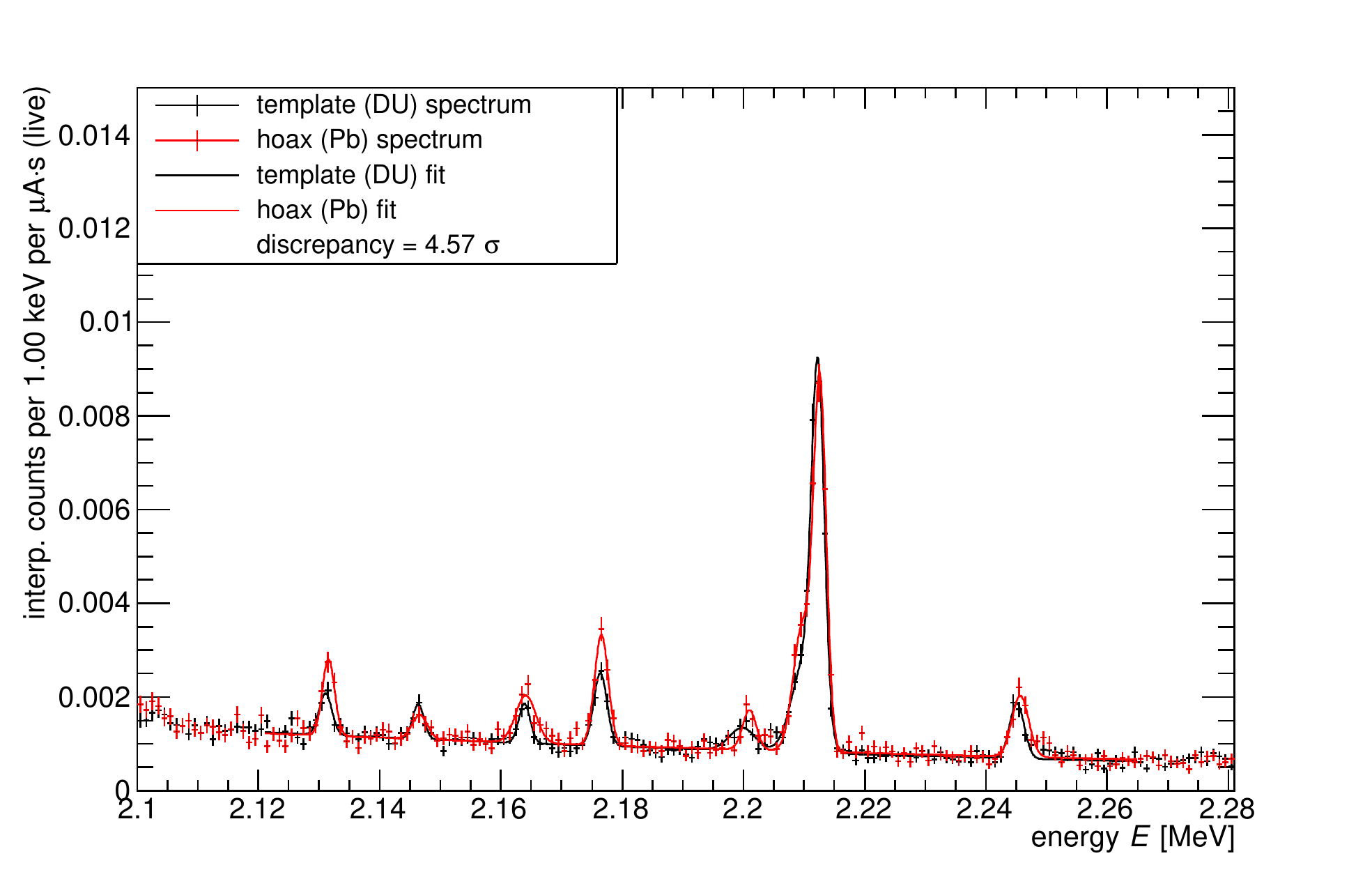}
    \caption{Summed spectra (points) and fits (curves) in the template II 
(black) vs hoax IId (red) verification measurement.}
    \label{fig:II_vs_IId}
\end{figure}

The interpolated histograms $\bar{r}$ (see Figs.~\ref{fig:I_vs_Ia}--\ref{fig:II_vs_IId}) are fit with an exponential background plus a series of Gaussian peaks as given in Eq.~\ref{eq:spectral_fit} of the main article. Such high-dimensional fits are achieved reliably by an iterative process: each peak is 
first fit individually with a five-parameter Gaussian plus linear background 
curve, where the initial parameter estimates and bounds for ROOT's $\chi^2$ 
minimization are computed using rough linear approximations. In the case of the 
closely-spaced $2.209$~MeV and $2.212$~MeV lines from U-238 and Al-27, 
respectively, an eight-parameter doublet fit is used instead. A first-order 
estimate of the continuum is then made by fitting the entire spectrum (including 
the peaks) with a single exponential curve. The two parameters of this 
exponential, along with the area, mean, and standard deviation estimates of each 
of the six peaks, are then input directly as starting estimates for the full 
26-parameter fit. Parameter bounds are established in a similar fashion by 
allowing some tolerance around the starting estimates.

\begin{sidewaystable*}[hbt]
\footnotesize
\centering
\caption{Proxy warhead verification measurements}\label{tab:configs_full}
\makebox[\textwidth][c]{
\begin{tabular}{@{\vrule height 10.5pt depth4pt  width0pt}cccccccc}
\hline\vspace{-5pt}
meas. & warhead & summed live charge & Al-27 fit peak rate & U-238 fit peak 
sum & Al-27 discrepancy & U-238 discrepancy & figure\\
  &  & $Q_\ell$ [{\textmu}A$\cdot \text{h\, (live)}$] & 
[({\textmu}A$\cdot$h)$^{-1}$] & $T$ [({\textmu}A$\cdot$h)$^{-1}$] & $\nu$ (vs 
template) & $\nu$ (vs template)\\
\hline
0 & template I & 50.0 & 99.1 $\pm$ 4.5 & 87.9 $\pm$ 4.6 & - & - & -\\
1 & hoax Ia (100\% Pb) & 67.5 & 98.7 $\pm$ 6.5 & 140.8 $\pm$ 4.8 & 
$-$0.051~$\sigma$ & 7.9~$\sigma$ & \ref{fig:I_vs_Ia}\\
2 & genuine candidate Ig & 61.0 & 105.3 $\pm$ 6.8 & 99.4 $\pm$ 4.8 & 
0.76~$\sigma$ & 1.7~$\sigma$ & \ref{fig:I_vs_Ig}\\
3 & hoax Ib (100\% Pb) & 54.8 & 112.0 $\pm$ 6.0 & 153.4 $\pm$ 4.8 & 1.7~$\sigma$ 
& 9.8~$\sigma$ & \ref{fig:I_vs_Ib}\\
\hline
4 & template II & 53.7 & 69.9 $\pm$ 7.5 & 59.0 $\pm$ 3.4 & - & - & -\\
5 & hoax IIc (100\% Pb) & 42.9 & 72.7 $\pm$ 8.4 & 109.9 $\pm$ 3.4 & 
0.25~$\sigma$ & 10.7~$\sigma$ & \ref{fig:II_vs_IIc}\\
6 & hoax IId (50\% Pb) & 73.7 & 85.8 $\pm$ 3.6 & 84.6 $\pm$ 4.5 & 1.9~$\sigma$ & 
4.6~$\sigma$ & \ref{fig:II_vs_IId}\\
\hline
\end{tabular}
}
\\
\footnotesize{Live charge, Al-27 fit rate, and U-238 fit rate are summed 
over three detectors. U-238 fit rate is additionally summed over six peaks. Uncertainties are $\pm$1~SD.}
\end{sidewaystable*}

\section{Extrapolation calculations for realistic warhead measurements}
\label{sec:si_extrap}

The extrapolations to more realistic warhead models and future dedicated 
verification systems are computed using Eq.~\ref{eq:d2ndEdOmega}, which predicts 
the detected count rate of a single NRF line for a given encryption foil and 
warhead geometry. All calculations assume isotropic NRF emission for simplicity 
(especially when spin states are unknown in e.g.~U-235), and use the same Pb 
filter transmission probability function $P_f(E) \sim 0.25$, intrinsic peak 
detector efficiency $\epsilon_\text{int} \simeq 0.16$, and single-detector 
geometric efficiency $\Omega_d/4\pi \simeq 1.3\times 10^{-3}$ used in this 
work's experiments. The incident bremsstrahlung spectrum $\phi_0(E)$ is computed 
in a Geant4 simulation of the 126~{\textmu}m Au radiator with an electron beam 
energy of 2.521~MeV as determined in Fig.~\ref{fig:endpoint}, and it is assumed 
that the simulated flux accurately predicts the flux that would be observed in 
the laboratory. For use in Eq.~\ref{eq:d2ndEdOmega}, $\phi_0(E)$ 
is approximated as a pencil beam impinging on the axis of a concentric-shell 
warhead. Calculations for different warhead models assume 
different foil compositions (though maintain the $X=3.18$~mm thickness) and 
therefore consider different NRF lines depending on whether a uranium or 
plutonium component would need to be verified. In both cases, the NRF line used 
in Eq.~\ref{eq:d2ndEdOmega} is chosen as the ground-state transition with the 
highest integrated cross section based on the values in Table~I 
of~\cite{ref:bertozzi_full}, excluding those (in particular, the Pu-239 
2143.56~keV transition) with poorly-understood nuclear level schemes. In all 
scenarios, a Pb hoax is constructed by replacing a weapon-isotope component with 
the same areal density of Pb. The runtime (quantified in `detector live microamp 
hours', i.e.~the triple product of the number of detectors, the beam current, 
and the live time) required to distinguish the NRF count rates $r_1$ and $r_2$ 
of the genuine and hoax warheads at a confidence $\nu'$ of $5\,\sigma$ is 
computed as
\begin{align}
    \nu' = \frac{r_2 - r_1}{\sqrt{r_1/C_\ell + r_2/C_\ell}}.
\end{align}
where the triple product $C_\ell$ is assumed to be equal in the two 
measurements. Table~\ref{tab:extrap_params} in the article lists the results of these 
calculations for six different warhead geometries. The extrapolation 
calculations given in the article, for instance, use 30 detectors and a 5~mA 
current to arrive at a measurement time of roughly 20 minutes (per object) for 
the second entry of Table~\ref{tab:extrap_params}. These 
calculations use only the counts in a single NRF line, but summing the lines 
from a single isotope will increase the statistics and reduce the quoted 
required measurement times.

We note that for more realistic (i.e.~non-slab) geometries, the 
precise alignment of the beam, warhead, foil, and detectors may become important 
for the prediction of absolute NRF count rates in a verification measurement. 
These systematic factors will cancel in the verification measurement, however, if 
they are kept constant between template and candidate measurements. For completeness, we estimate 
here the effect of a misalignment of the warhead transverse to the beam, which 
effectively changes the warhead thickness $X$. If we take for simplicity a nominal
spherical shell of DU with inner and outer radii of 6.3 and 6.5~cm, 
respectively, and measure until 2000 counts are obtained in the 2.176~MeV line 
of U-238, we find (using Eq.~\ref{eq:d2ndEdOmega}) that a $1\,\sigma$ 
discrepancy in the observed 2.176~MeV rate requires a misalignment of 
approximately $1.6$~cm. Such displacements may also affect the solid angle 
integration of Eq.~\ref{eq:d2ndEdOmega}, changing the geometric efficiency by 
factors on the order of $10\%$, depending on the foil-to-detector distance. 
However, geometric control on scales $<1$~cm is feasible with appropriate survey 
equipment.  This effect is further mitigated by the fact that the bremsstrahlung
beam has a spatial width and thus samples an extended area of warhead at a given time.  For these
experiments, the opening half-angle of the cone of beam was approximately $5^\circ$ and the photon illumination
in this cone was relatively uniform (to better than 10\%). The collimation of the beam could be adjusted
to further cancel misalignment effects by adjusting the size and uniformity of the beam spot on the 
inspected object. Finally, manufacturing tolerances in the warheads or true 
warhead-to-warhead variation in component sizes may also affect the results of 
the template measurement, but estimates of such variations are not available in 
the open literature.

\begin{figure}[!htb]
    \centering
    \includegraphics[width=\columnwidth]{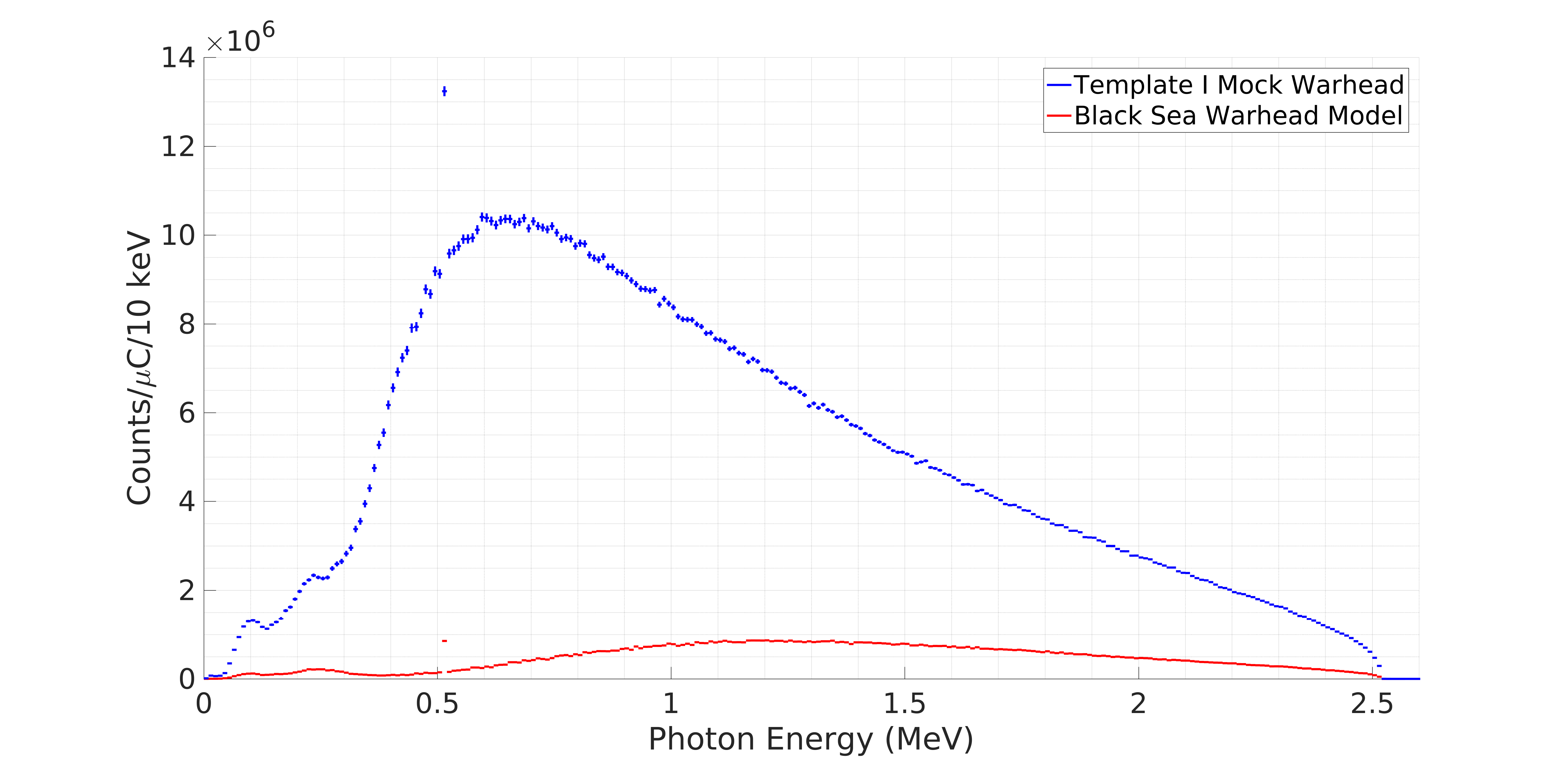}
    \caption{Transmitted spectrum of the bremsstrahlung beam at the encryption foil
    for the template I mock warhead and the Black Sea model warhead \cite{ref:fetter1990gamma}.}
    \label{fig:foiltrans}
\end{figure}

An additional concern regarding extrapolation to measurements
at mA-scale beam currents is the capability of the HPGe detectors to handle
the event rate increase and the additional loss of live time to
pile-up events.  While a 5~mA current will produce 200$\times$ as many bremsstrahlung
photons relative to the 25~{\textmu}A currents of the experiments described in this
article, the rates in the HPGe detectors will increase by a lesser factor.  Realistic
inspection objects will be larger than the mock warheads used for this work and thus will prevent a greater fraction of the
beam from reaching the encryption foil.  Since the event rate in the HPGe detectors is dominated
by photons scattered from the foil, this attenuation reduces the event rate in the detectors.
To estimate the size of this effect the transmission of the bremsstrahlung beam through the warhead
test object to the encryption foil was simulated for two scenarios --- the ``template I'' mock warhead
and the WGPu+DU ``Black Sea'' warhead model consisting of spherical shells of WGPu, high explosives,
and a uranium tamper \cite{ref:fetter1990gamma}.  Fig. \ref{fig:foiltrans} shows the transmitted
spectrum through each of these objects per 1~{\textmu}C of electron beam on target.  The total transmitted
photon rate through the mock warhead is approximately 11 times higher than that of the more realistic
Black Sea model, and the rate at the high end of the spectrum ($\gtrsim$2 MeV) is approximately 6 times
higher for the mock warhead.  Since photons are more likely to eventually cause events in the detectors
if they strike the foil at high energies, the latter factor of $\sim$6 is taken as a conservative estimate
of the rate reduction due to the thicker warhead.  Thus, the increase in the event rate in the detectors
between the experiments described in this work and a realistic warhead under inspection with a 5~mA electron
beam current will be approximately 30.  At this rate, the dead time fraction due to pile-up events would
be $\sim$60\% while the increase in the fixed event rate processing would result in a total dead time
fraction of $\sim$90\%.  This corresponds to a live time reduction of approximately an order of magnitude
relative to the mock warhead experiments.  This may be mitigated in a future realistic verification
scenario by increasing the number of detectors to directly increase the live time, and in the future it is
likely that HPGe detectors capable of operating at MHz-scale rates will be available that would be more
than sufficient for this application \cite{ref:pnnlhpge}.  Additionally, the 5~mA beam current assumption
here is merely a starting point to provide a reference for estimated measurement times and may be optimized
to achieve a balance between detector rates and measurement times.  As also noted in the text, minimal effort
was made to optimze the low-energy photon filters in front of the HPGe detectors for this experiment and it
is likely that further rate reductions could be achieved by optimized shielding.

\section{Multi-line inference}
The equation for the predicted NRF rates (Eq.~\ref{eq:d2ndEdOmega} or its 
integrated form), contains multiple quantities that are kept secret from the 
inspector, and thus cannot be used alone to infer the warhead thickness~$D$ from 
a physically-encrypted spectrum. However, it is possible to develop a system of 
equations from Eq.~\ref{eq:d2ndEdOmega}---one equation per NRF peak---in which 
case there may be at least as many equations as unknowns and inference may be 
possible. This technique may be especially straightforward to realize by using 
two NRF lines of the same isotope with similar resonance energies $E_r$, such as 
the 2.176~MeV and 2.209~MeV U-238 lines (if the latter were not obscured by the 
the 2.212~MeV Al-27 line) in any of the above NRF spectra. Taking a ratio of 
Eq.~\ref{eq:d2ndEdOmega} for the two lines allows one to cancel systematic 
factors such as the $\epsilon_\text{int}(E')$, $P_f(E')$, and number densities 
$N$, and to approximately cancel slowly varying functions of energy $E$ such as 
$\phi_0(E)$ or perhaps even $\phi_t(E)$. If the $\phi_t(E)$ are canceled, the 
ratio of observed counts may then be used to estimate the foil thickness~$X$, 
which could in turn be used to estimate~$D$; more complicated procedures are 
required if only the $\phi_0(E)$ are canceled. This information security question may be solved by use of the encryption plates, which obscure the true value of $D$ and permit inference only on some upper bound $D + \Delta D$.


\end{document}